\documentclass[11pt,a4paper,amssymb,amsmath, tightenlines]{article}

\usepackage{amsmath, amssymb, amsthm, latexsym, mathrsfs, url, color, epsfig, graphics, float, setspace}
\usepackage{sectsty}
\usepackage{caption,subcaption}

\usepackage[T1]{fontenc}
\usepackage[bitstream-charter]{mathdesign}
\usepackage[bottom]{footmisc}

          		 \usepackage{bbm}
\usepackage{geometry,enumerate}
\usepackage{bbm}

\usepackage{graphicx}				
	\def\beq{\begin{equation}}
\def\eeq{\end{equation}}
\newcommand{\barr}{\begin{array}}
\newcommand{\earr}{\end{array}}
\newcommand{\bqr}{\begin{eqnarray}}
\newcommand{\bqq}{\begin{eqnarray*}}
\newcommand{\eqr}{\end{eqnarray}}
\newcommand{\eqq}{\end{eqnarray*}}
\newcommand{\rd}{\textrm{d}}
\newcommand{\e}{\textrm{e}}

\newcommand{\E}{\mathbb{E}}

\newcommand{\RR}{\mathbb{R}}				
\numberwithin{equation}{section}

\theoremstyle{plain}
\newtheorem{thm}{Theorem}[section] 

\theoremstyle{definition}
\newtheorem{defn}[thm]{Definition} 

\newtheorem{prop}[thm]{Proposition} 

\usepackage{doc}
\usepackage{url}
\usepackage{graphicx}
\usepackage{epstopdf}
\title{Simultaneous Trading in `Lit' and Dark Pools
}
\author{M. Alessandra Crisafi$^{\ast}$ and Andrea Macrina$^{\ast\, \dagger}$ \\ 
$^{\ast}$Department of Mathematics, University College London \\ Gower Street, London WC1E 6BT, U. K.\\
$^{\dagger}$Department of Actuarial Science, University of Cape Town \\ Rondebosch 7701, South Africa }

\begin{document}

\maketitle
\vspace{-1cm}
\begin{abstract}
\noindent We consider an optimal  trading problem over a finite period of time during which an investor has access to both  a standard exchange and a dark pool. We take the exchange to be an order-driven market and  propose a continuous-time setup for the best bid price and the market spread, both modelled by L\'evy processes. Effects on the best bid price arising from the arrival of limit buy orders at more favourable prices, the incoming market sell orders potentially walking the book, and deriving from the cancellations of limit sell orders at the best ask price are incorporated in the proposed price dynamics. A permanent impact that occurs when `lit' pool trades cannot be avoided is built in, and an instantaneous impact that models the slippage, to which all lit exchange trades are subject, is also considered. We assume that the trading price in the dark pool is the mid-price and that no fees are due for posting orders. We allow for partial trade executions in the dark pool, and we find the optimal trading strategy in both venues. Since the mid-price is taken from the exchange, the dynamics of the limit order book also affects the optimal allocation of shares in the dark pool. We propose a general objective function and we show that, subject to suitable technical conditions, the value function can be characterised by the unique continuous viscosity solution  to the associated  partial integro differential equation. We present two explicit examples of the price and the spread models, and derive the associated optimal trading strategy numerically. We discuss the various degrees of the agent's risk aversion and further show that roundtrips---i.e. posting the remaining inventory in the dark pool at every point in time---are not necessarily beneficial. 

\vspace{.25cm}

\noindent
\end{abstract}
\setlength{\topmargin}{0in}

\section{Introduction}
\label{intro}

The focus in this paper is put on trading in `lit' and dark pools. We study how an investor's optimal trading programme is affected by the availability of a dark pool as an alternative venue with its own set rules. Since the trade execution price in the dark pool is related to the best bid and ask prices in the standard exchange (often referred to as the `lit' pool or `lit' exchange), we propose at the same time novel microstructure models for the price dynamics in the lit exchange. In this paper we assume the exchange to be an order-driven market and put forward continuous-time models for the best bid price and for the bid-ask spread. The mid-price is of course recovered by adding half the spread to the best bid price. We capture the permanent price impact by market activity in the drift function and in the Poisson random measures. The spread process is constructed similarly so as to obtain the same structure for the  best ask price when compared to the best bid price process. Optimal trading in LOBs, of which microstructural features are modelled, is for example also treated in (i) Cartea and Jaimungal  \cite{c1}  who model the spot price via a hidden Markov chain to capture the switches between price regimes, (ii) Cartea et al.  \cite{77}  who model the deviation of mid-price from its long-term mean via a jump-diffusion process, and (iii) Obizhaeva and Wang \cite{24}, and Alfonsi et al. \cite{1} who propose trading models by taking into account shape functions for the LOB. Research in optimal execution in an order-driven market has its roots in the papers by Bertsimas and Lo \cite{3}, and Almgren and Chriss \cite{2}. More recent contributions include, e.g.,  Pemy and Zhang \cite{25},  Pemy et al. \cite{26}, Gatheral and Schied \cite{15}, Brigo and Di Graziano \cite{5}, Moazeni et al. \cite{23}, Cartea et al. \cite{c2}, and Ishitani and Kato \cite{20}. Cartea and Jaimungal \cite{7} consider a continuous-time, jump-diffusion mid-price model and explicitly take into account the impact of the market activity on the mid-price. In the same context, we also mention the works by Gu\'eant et al. \cite{a1} and Guilbaud  et Pham \cite{a2}, who treat optimal liquidation via limit orders. 
A dark pool is an alternative trading venue where buy and sell orders are not publicly displayed so that the participants' identity is not revealed.  Other advantages of trading in a dark pool include (i) the possibility of submitting large orders without impacting the market price, and (ii) trading at  more favourable conditions---the dark pool trading price lies between the best bid and the best ask prices quoted in the standard exchange. (For simplicity, we will assume that the dark pool utilises the standard exchange mid-price.) On the other hand, dark pools do not guarantee execution of the trade for it is subject to the availability of a trade counterparty. Dark pools might also reduce market liquidity since large and institutional investors, in order to protect private information, resort to these alternative trading venues where orders remain hidden to the general market participants. Most of the research effort in dark pools has been devoted to  the impact of dark pools on market welfare. A game-theoretic approach has often been used to model the two competing trading venues, so to find equilibrium prices and optimal strategies of the market players, also including the effect of private information  and adverse selection on trading. This branch of research includes the early work by Hendershott and Mendelson \cite{18}, who extend the Kyle \cite{22} model to capture the dynamics of the interplay between investors, dark pool and standard exchange, followed by Degryse et al. \cite{13}, Buti et al. \cite{6} and  Dani\"{e}ls et al. \cite{12}. The work by Ye \cite{30} and Zhu \cite{32} focus on the effects on price discovery due to the migration of order flows from the exchanges to the dark pools, when unobservable liquidity is added to the model.

Kratz and Sch\"{o}neborn \cite{21} consider trading in the dark pool within the classical field of optimal liquidation. They model the LOB mid-price by an exogenous square-integrable martingale and regard the dark pool as a complete-or-zero-execution venue where the arrival of trading counterparties follows a Poisson process.  In this context, we also refer to work by Horst and Naujokat \cite{H1}, in which the authors find the optimal  strategy when trading in an illiquid market. The agent under consideration seeks to minimise the deviation from a given target while submitting market orders in the standard exchange and ``passive orders'' in a dark pool. The execution of  dark-pool orders is modelled via Poisson processes. As we shall see, we incorporate such a feature in the model presented in Section \ref{eq:dp}.

The goal in this paper is to find the optimal trading strategy when utilising standard exchange and the services of a dark pool. Since the smoothness of the value function is not guaranteed on the whole domain, there are no classical solutions to the optimisation problem, in general. Therefore we resort to a weaker notion, the viscosity solutions, and we show that the value function is the unique viscosity solution of the Hamilton-Jacobi-Bellman partial integro differential equation (HJB PIDE) associated with the optimisation problem. There is a vast literature on the theory of viscosity solution of HJB equations of which state variables  are driven by general Markov processes. We here refer in particular to Pham \cite{27,28}, and Fleming and Soner \cite{14} for technical details. Our results concerning the exchange are coherent with the ones obtained by Chevalier et al. \cite{8}. Indeed we find that the optimal strategy depends on the book's dynamics. Furthermore, we show that this is also the case in the presence of a dark pool.

The paper is organised as follows: after the introduction, in  Section \ref{eq:exchange} we discuss the optimal trading problem and present our LOB model. Then we formulate the optimisation problem when only standard exchanges are available to an investor. We consider a generalised objective function and derive the HJB PIDE for the control problem. In Section \ref{eq:dp}, we introduce the dark pool by modifying the setup presented in the previous section. We  propose an objective function that accounts for trading in the alternative pool and obtain the associated  HJB PIDE. In Section \ref{eq:numerics}, we provide two numerical examples of the optimal strategy and we focus on the roles played by the parameters of the explicit model. We conclude with Section \ref{Sect.Conc} by summarising the contributions offered in this paper. In order for the present work to be self-contained, we collect the mathematical conditions, the propositions and the related proofs at the basis of the viscosity property of the value function in the appendix.

\section{Price model and optimal trading in a standard exchange\label{eq:exchange}}

We take the view of an investor who seeks to liquidate a sizeable amount of shares over a finite period of time $[t,T]$, where the initial time $t$ lies in the interval $[0,T)$. As far as an optimal liquidation strategy is concerned---which in our setup involves the submission of market sell orders only---it suffices  to model the best bid (or the best ask in the case of an optimal acquisition), since  we  consider a temporary price impact to account for the slippage effect. In the short run, the best bid price at time $u\in[t,T]$  depends on its initial state $s^b$ at time $t$ and on the cumulated market activity up to time $u$. 

We fix a probability space $(\Omega,\mathcal F,\mathbb P)$ equipped with a filtration $\{\mathcal F_u\}_{{0}\leq u\leq T}$ satisfying the usual conditions. In what follows, we assume that the task is to liquidate an amount $X(t)=x\in[0,X]\subset \RR_{+}\cup0$ of shares and write for the inventory dynamics 
\begin{equation}
\rd X(u)=-\nu\,(u)\rd u,
\label{eq:inventory}
\end{equation}
where the rate of trading $\nu:=\{\nu\,(u)\}_{t\leq u\leq T}$ is  the control process of the stochastic optimisation problem presented in Section \ref{eq:optim}. We model the best bid price process by
\begin{equation}
\rd S^b\left(u\right)=\mu^b\left(u,S^b({u}),\nu\,(u)\right)\rd u+\sum_{i=1}^2\int_\RR\!\!h^{b}_i\left(u,S^b(u^{-}),z_i\right)\tilde{\Gamma}_i^b\left(\rd u,\rd z_i\right)\!,\!
\label{eq:bid}
\end{equation}
and the spread process by
\begin{equation}
\begin{split}
\rd  \Delta\left(u\right)=\mu\left(u,\Delta(u),\nu\,(u)\right)\rd u&+\sum_{i=1}^2\int_\RR\!\!h_i\left(u,\Delta(u^{-}),y_i\right)\tilde{\Gamma}_i\left(\rd u,\rd y_i\right)\\
&+\sum_{i=1}^2\int_\RR\!\!h^{b,\Delta}_i\left(u,S^b(u^{-}),\Delta(u^{-}),z_i\right)\tilde{\Gamma}_i^b\left(\rd u,\rd z_i\right)\!,\\
\label{eq:ask}
\end{split}
\end{equation}
with  initial values $S^{b}(t)=s^{b}$ and $\Delta(t)=\Delta$. For $i=1,2$,  $\Gamma^{b}_i\big(\rd u,\rd z_i\big)$ and $\Gamma_i\big(\rd u,\rd y_i\big)$ are independent and homogeneous Poisson random measures. Their compensated versions are defined by  $\tilde{\Gamma}_i\big(\rd u,\rd y_i\big)=\Gamma_i\big(\rd u,\rd y_i\big)-\rd u\times \Pi_i\big(\rd y_i\big)$ and by $\tilde{\Gamma}^{b}_i\big(\rd u,\rd z_i\big)=\Gamma^{b}_i\big(\rd u,\rd z_i\big)-\rd u\times \Pi^{b}_i\big(\rd z_i\big)$, where  $\Pi_i\big(\rd y_i\big)$ and $\Pi^b_i\big(\rd z_i\big)$ are $\sigma$-finite L\'evy measures on $\RR$ such that
$$
\int_{|z_i|\geq 1}\Pi_i^b\big(\rd z_i\big)<+\infty,\quad \int_{|{y}_i|\geq 1}\Pi_i\big(\rd y_i\big)<+\infty.
$$
We refer to Gihman and Skorohod \cite{16} for a technical account on Poisson random measures. 

We say that $\nu$ is {\itshape admissible} if (i) it is progressively measurable, (ii) it is such that \linebreak
\noindent $\E\left[\int_t^T\big|\mu\left(u,\Delta(u),\nu\,(u)\right)\big|^2+\big|\mu^b\left(u,S^b(u),\nu\,(u)\right)\big|^2\rd u\right]<\infty$, and (iii) it belongs to a compact set $\mathcal V=[0,N]\subset[0,\infty)$. The functions $\mu$, $\mu^{b}$, $h_{i}$, $h^{b}_{i}$ and $h^{b,\Delta}_{i}$ can be chosen such that both the best bid price and the spread are non-negative, and in ways to reproduce market features. In Equation (\ref{eq:bid}) we consider positive jumps which model the incoming limit buy orders at a more favourable price, and negative jumps to account for cancellations of limit buy orders and market sell orders which walk the LOB. Explicit examples will be provided in Section \ref{eq:numerics}. Equation (\ref{eq:ask}) models the spread process, of which width is affected by both, the market activity on the ask side and on the bid side of the book. In particular, when the best bid price increases, the spread process should decrease by the same amount, assuming that the ask price remains unchanged. We therefore include the jump processes of the bid price in the dynamics of the spread process through the functions $h^{b,\Delta}_{i}$. The activity on the ask side is accounted for by the functions  $h_{i}$. We further include the trading rate $\nu\,(u)$ in the drift of the spread process. This is financially justified for $\nu\,(u)$ has an impact on the best bid price---which decreases---and thus it has an opposite effect on the spread. The best ask price,
\begin{equation}\label{eq:SaSm}
S^a(u)=S^b(u)+\Delta(u),
\end{equation}
shares the same model structure as the best bid price given that the spread process $\{\Delta(u)\}$ has the form (\ref{eq:ask}). Although we view the optimisation problem from the perspective of liquidating orders,  the model proposed via Equations (\ref{eq:inventory}), (\ref{eq:bid}), and (\ref{eq:ask}) can be  adapted to the case of optimal acquisition.  
Last, we define the cash (wealth) process by
$$
\rd W(u)=\mu^{w}(u,S^{b}(u),\nu\,(u))\rd u,
$$
with initial cash $W(t)=w$. The function $\mu^{w}$ models the instantaneous gains made by the investor through selling shares, possibly taking into account the temporary price impact of trades.

\subsection{Optimal control problem and viscosity solution \label{eq:optim}}
An investor seeks to liquidate $x$ shares over a finite period of time $[t,T]$. The goal is  to maximise the total reward, while minimising the unfavourable impact of executed trades on the best bid price $\{S^b(u)\}$. We assume that the investor trades by means of market sell orders only, and we introduce a stopping time $\tau$, 
 \begin{equation}
 \tau:=\inf\,\{u\geq t\,\vert \,X(u)\leq0\}\wedge T,
 \label{eq:tau}
 \end{equation}
that describes the first time the inventory is depleted, should such an event occur before the terminal date $T$. For notational simplicity, in this section we introduce the space $\mathcal O=[0,X]\times \RR_{+}\times \RR$ and let the vector of the state variables be defined by  $\boldsymbol{X}_{t,\,\boldsymbol{x}}(u)=\big(X_{t,\,x}(u),S^b_{t,\,s^b}(u),W_{t,\,w}(u)\big)\in\mathcal O$, with initial values ${\boldsymbol{x}}=(x,s^b,w)\in\mathcal O$.   We propose a general objective function of the form
\begin{equation}
\begin{split}
\!\!V\left(t,\boldsymbol{x}\right)\!=\!\sup_{\nu \in\mathcal V}\E&\!\left[\int_{t}^\tau\!\!\! \e^{-r(u-t)}f\!\left(u,\boldsymbol{X}_{t,\,\boldsymbol{x}}(u),\nu\,(u)\right)\rd u+\ \e^{-r(\tau-t)}g\!\left(\boldsymbol{X}_{t,\,\boldsymbol{x}}(\tau)\right)\right], \!\!
\end{split}
\label{eq:valuefunction}
\end{equation}
where $r\geq0$ is a discount rate and $\E[\cdot]$ is the  expectation given the initial state of the system $(t,\boldsymbol{x})\in[0,T)\times\mathcal O$. We interpret the discount factor as the urgency of the agent to liquidate and enter in possession of the revenues deriving from the sales. That is, even a risk-neutral investor ($\gamma=0$) may prefer to liquidate faster as a result of his preference for a more immediate liquidity. The function $f:[0,T]\times\mathcal O\times\mathcal V\rightarrow \RR$ may have several interpretations. For example it may represent the gains made from the shares sale, or it may be the return obtained by following a mean-variance investment strategy. The function $g:\mathcal O\rightarrow \RR$ may represent the terminal reward obtained by a block trade liquidation of the remaining inventory at time $T$. However, $g$ may also be a penalty resulting from failing to liquidate the whole inventory by time $T$. We let $\boldsymbol{ p}:=(p_1,p_2,p_{3})\in\RR^3$, and we define the operator ${\mathcal H}$ by
\begin{equation}
{\mathcal H}\left(t,\boldsymbol{x},\boldsymbol{ p}\right)=\sup_{v\in\mathcal V}\Big\{f\left(t,\boldsymbol{x},v\right)\!-vp_1+\mu^b\left(t,s^b,v\right)p_2+\mu^{w}(t,s^{b},v)p_{3}\Big\}.
\end{equation}
Also, for $0\leq\delta\leq1$ and $\varphi\in{PB}$, where ${PB}$ is the space of continuous and polinomially-bounded  functions on $[0,T]\times\mathcal O$, we let the operator $\mathcal B^\delta_{b}$ be defined  by
\begin{equation}
\mathcal B^\delta_{b}\left(t,\boldsymbol{x},p_2,\varphi\right)\!=\!\!\sum_{i=1}^{2}\int_{{|z_{i}|\geq\delta}}\!\left[\varphi\left(t,x,s^b\!+ h^{b}_i\left(t,s^b\!,z_{i}\right),w\right)-\varphi\left(t,\boldsymbol{x}\right)- h^{b}_i\left(t,s^b\!,z_{i}\right)p_2\right]\!\Pi^{{b}}_{i}\left(\rd z_{i}\right),
\label{eq:integral}
\end{equation}
and, for $\varphi\in\mathcal C^{1,1}([0,T]\times\mathcal O)$, we let $\mathcal B^{b}_\delta$ be defined  by
\begin{equation}
\begin{split}
\mathcal B^{b}_\delta\left(t,\boldsymbol{x},\varphi\right)\!=\!\!\sum_{i=1}^{2}\int_{{|z_{i}|\leq\delta}}&\!\!\left[\varphi\left(t,x,s^b\!\!+ h^{b}_i\!\left(t,s^b,z_{i}\right),w\right)-\varphi\left(t,\boldsymbol{x}\right)\!-\! h^{b}_i\!\left(t,s^b,z_{i}\right)\!\!\frac{\partial\varphi}{\partial{s^{b}}}\left(t,\boldsymbol{x}\right)\!\right]\!\Pi^{{b}}_{i}\left(\rd z_{i}\right).
\label{eq:integral1}
\end{split}
\end{equation}
Furthermore, we define
\begin{equation}
\mathcal B_{b}\left(t,\boldsymbol{x},\varphi\right)=\mathcal B_{b}^\delta\left(t,\boldsymbol{x},\frac{\partial\varphi}{\partial{s^{b}}},\varphi\right)+\mathcal B^{b}_\delta\left(t,\boldsymbol{x},\varphi\right).
\label{eq:welldefn}
\end{equation}
Assumption (\ref{eq:hhhbbb}) in the appendix ensures that the integrals in (\ref{eq:integral}) and (\ref{eq:integral1}) are well defined, and similarly for (\ref{eq:welldefn}). Details on their convergence and uniform boundedness can be found in, e.g., Pham \cite{27}.
\noindent Standard arguments from dynamic programming (DP) suggest that the value function of the optimal control problem  (\ref{eq:valuefunction}) satisfies the following HJB PIDE:
\begin{equation}
\begin{split}
&rV\left(t,\boldsymbol{x}\right)-\frac{\partial V}{\partial t}\left(t,\boldsymbol{x}\right)-\mathcal H\left(t,\boldsymbol{x},D_{\boldsymbol{x}}V\right)-\mathcal B_{b}\left(t,\boldsymbol{x},V\right)=0,
\label{eq:pde}
\end{split}
\end{equation}
on $[0,T)\times\mathcal O$, with terminal condition $V(T,\boldsymbol{x})=g(\boldsymbol{x})$.  Since one cannot guarantee the smoothness of $V(t,\boldsymbol{ x})$ on the whole domain, one cannot discuss the solution of the HJB PIDE in the classical sense. We resort to the weaker notion of viscosity solutions and  borrow the definition from Pham \cite{27}.
\begin{defn}
{\itshape A continuous function $V:[0,T)\times\mathcal O\rightarrow\RR$ is a viscosity subsolution (resp. supersolution) of the HJB Equation (\ref{eq:pde}) if
\begin{equation}
\begin{split}
r\phi\left(\bar{t},\boldsymbol{\bar x}\right)-\frac{\partial \phi}{\partial t}\left(\bar{t},\boldsymbol{\bar x}\right)-\mathcal H\left(\bar{t},\boldsymbol{\bar x},D_{\boldsymbol{ x}}\phi\right)-\mathcal B_{b}\left(\bar{t},\boldsymbol{\bar x},\phi\right)&\leq0,
\end{split}
\end{equation}
(resp. $\geq0$), for any $\phi\in \mathcal C^{1,1}([0,T]\times\mathcal O)\cap PB$ such that $V(t,\boldsymbol{ x})-\phi(t,\boldsymbol{x})$ attains its maximum (resp. minimum) at $(\bar{t},\boldsymbol{\bar x})\subset[0,T)\times\mathcal O$. A continuous function is a viscosity solution if it is both a viscosity subsolution and a viscosity supersolution.}
\end{defn}
In the appendix, we provide the characterisation of the value function $V$ by the unique continuous viscosity solution of (\ref{eq:pde}).

The optimal trading strategy by Bertsimas and Lo \cite{3}, whereby one liquidates at a constant rate $\nu\,(u)=X(u)/(T-u+1)$, is recovered within this setup as follows: (i) consider profit maximisation from trading for a risk-neutral investor as the objective function, and (ii) require the bid price process (\ref{eq:bid}) be a (local) martingale. Under such assumptions, the optimal trading strategy is independent of the LOB dynamics (cfr. Figure  \ref{eq:lit2}, left). By contrast, the solutions to the concrete examples of the liquidation problems treated in the next section confirm the view---also in agreement with the work by Chevalier et al. \cite{8}---that the book dynamics is crucial when dealing with optimal execution. Further details can be found in Section \ref{eq:numerics} (Figures \ref{eq:lit1}, \ref{eq:lit2} and \ref{eq:lit3}).

\section{Dark pool \label{eq:dp}}
We introduce the possibility for the investor to trade in a standard exchange and in a dark pool at the same time. The investor faces  a trade-off between a costly and  sure trade  execution in the exchange and the alternative of a more remunerative but uncertain execution in the dark pool.  

We take this opportunity to emphasise the differences between the recent literature---e.g.  Kratz and Sch\"{o}neborn \cite{21}, and Horst and Naujokat \cite{H1}---and the present work. Firstly, we model both the best bid price and the spread process, thus allowing for a more realistic description of market features, for there is no such thing as a mid-price in the market. Furthermore, by modelling both the bid price and the spread process, we also include the effect of the sell side of the LOB on the optimal strategy, which is justified since the two sides of the LOB are known to be highly correlated. We also include a permanent impact by the trades on the market prices. We do not require for the price processes to be martingales, as it is well known that they mean-revert to their long-term value. We further consider the possibility of partial execution in the dark pool, and we do not restrict the problem to full liquidation by $T$. Finally we propose a class of models that can be adapted to (i) market-specific features, e.g. mean-reversion, seasonality, and intra-day patterns of the price processes, as well as to (ii) agent-specific preferences of calculating the P\&L and/or measuring utility and risk-aversion. Such generalisations come at a cost and, in particular, we are not in the position to provide analytical trading strategies.
As specified earlier in this paper, we assume that the trades in the dark pool take place at the mid-price $\{S^m(u)\}$ given by
\begin{equation}
S^m(u):=S^b(u)+\tfrac{1}{2}\Delta(u)\,.
\label{eq:middp}
\end{equation}
\indent The  purpose of this section is to find the optimal trading strategy for each venue at any time $u\in[t,T)$. We denote the optimal order size in the  dark pool by  $\{\eta(u)\}$. Thus, the control process will be given by a  vector-valued process $\boldsymbol{\nu}=\{\boldsymbol{\nu}\,(u)\}_{t\leq u\leq T}$ defined by  $\boldsymbol{\nu}\,(u):=(\nu\,(u),\eta(u))\in\mathcal Z$, where $\{\nu\,(u)\}$ is progressively measurable, $\{\eta(u)\}$ is predictable and $\mathcal Z:=\mathcal V\times\mathcal N=[0,N]\times[0,N]\subset[0,\infty)^{2}$ is a compact set. We modify the inventory and the wealth dynamics since they now also depend on the dark pool activity. Along the lines of Horst and Naujokat \cite{H1} we model the dark-pool execution part by a jump process, but we choose a compound Poisson process  to account for partial execution of the submitted order. We thus write
\begin{equation}\label{inventdp}
\rd X(u)=-\nu\,(u)\rd u-\eta(u)\rd J(u),
\end{equation}
for the combined inventory and
\begin{equation}
\rd W(u)=\mu^{w}(u,S^{b}(u),\nu\,(u))\rd u+h^{w}(u,S^{b}(u^{-}),\Delta(u^{-}),\eta(u))\rd J(u)
 \label{eq:cash}
\end{equation}
for the wealth arising from simultaneously trading in the lit and the dark pools.
In the above, $J(u):=\sum_{i=1}^{N(u)}z^{w}_{i}$ is a compound Poisson process with intensity $\lambda^{w}$ and i.i.d. random variables $z^{w}_{i}$ supported on $[0,1]$, which model the executed portion of the submitted order. 
To simplify the notation, we introduce the space $\mathcal O:=[0,X]\times\RR^2_{+}\times \RR$ and we define a vector of state variables $\boldsymbol{X}_{t,\,\boldsymbol{x}}(u)=\big(X_{t,\,x}(u),S^b_{t,\,s^b}(u),\Delta_{t,\,\Delta}(u),W_{t,\,w}(u)\big)\in\mathcal O$ with initial values $\boldsymbol{x}=(x,s^b,\Delta,w)\in\mathcal O$. Next we consider a generalised optimisation problem of the form 
\begin{equation}
\begin{split}
V\!\left(t,\boldsymbol{x}\right)=\sup_{\boldsymbol{\nu}\in\mathcal Z}\E\!\left[\int_{t}^\tau\!\! \e^{-r(u-t)}f_1\!\left(u,\boldsymbol{ X}_{t,\,\boldsymbol{x}}(u),{\nu}(u)\right)\rd u+\e^{-r(\tau-t)}g_1\!\left(\boldsymbol{X}_{t,\,\boldsymbol{x}}(\tau)\right)\right] ,
\label{eq:valuefunction1}
\end{split}
\end{equation}
where $\tau$ is defined by (\ref{eq:tau}), and $\E[\cdot]$ is the  expectation given the initial state of the system $(t,\boldsymbol{x}\in[0,T]\times\mathcal O)$. The function $f_{1}$ may play the role of a running gain or penalty criterion. We include  the lit-pool trading rate $\nu$ in $f_{1}$ as it may reflect a penalisation for the information leakage to which  publicly-displayed orders are subject to. The function $g_{1}$ is the terminal reward function which includes the terminal cash and a penalty for the remainder of the inventory at time $\tau\wedge T$. 
The HJB PIDE for trading simultaneously in the lit and the dark pool is analogous to the one presented in Section \ref{eq:optim}. We skip the details at this stage, but write the exact formulation in the appendix for completion and ease of comparison with the one for trading optimally only in the standard exchange. We further note that one can recover the setting presented in Section \ref{eq:optim} by specifying $z^{w}_{i}\equiv0$. We emphasise here that Kratz and Sch\"{o}neborn \cite{21} model the dark pool as a complete-or-zero-execution venue, where the placed order is either fully executed, when a suitable counterparty is found, or it expires unexecuted. Their optimal strategy in the single-asset case---which consists of placing small portions of the inventory in the exchange and all the remainder in the dark pool---can be obtained as a special case  of the present model (see Figure  \ref{eq:krsh}). 

\section{Explicit price models and numerical computation of the trading strategy \label{eq:numerics}}
In this section we investigate in detail two particular examples of the best bid price and the spread processes defined by  Equations (\ref{eq:bid}) and (\ref{eq:ask}). In the numerical solutions that follow, the chosen parameters are for illustrative purposes. An analysis of real data is beyond the scope of the present work.

\subsection{Mean-reverting model \label{eq:meanr}}
\begin{figure}[H]
\centering
\includegraphics[width=300pt,height=180pt]{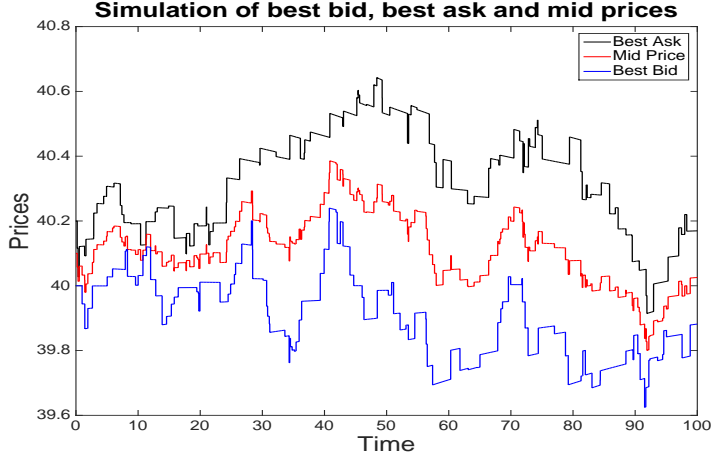}
\caption{\footnotesize{Simulation of the best ask, best bid and mid prices for a time frame of 100 seconds. We set $\lambda_1^b=\lambda_2^b=\lambda^{\Delta}_1=\lambda^{\Delta}_2=0.5$, $z_{i},y_{i}\sim U[0,0.1]$, $s^{b}=40$, $\Delta=0.2$, $\bar \Delta=0.1$, $\bar S=40.1$, $\kappa^{b}=\kappa^{\Delta}=2\times10^{-5}$.
}}
\label{eq:gr1}
\end{figure}

\begin{figure}[H]
\centering
\includegraphics[width=205pt,height=170pt]{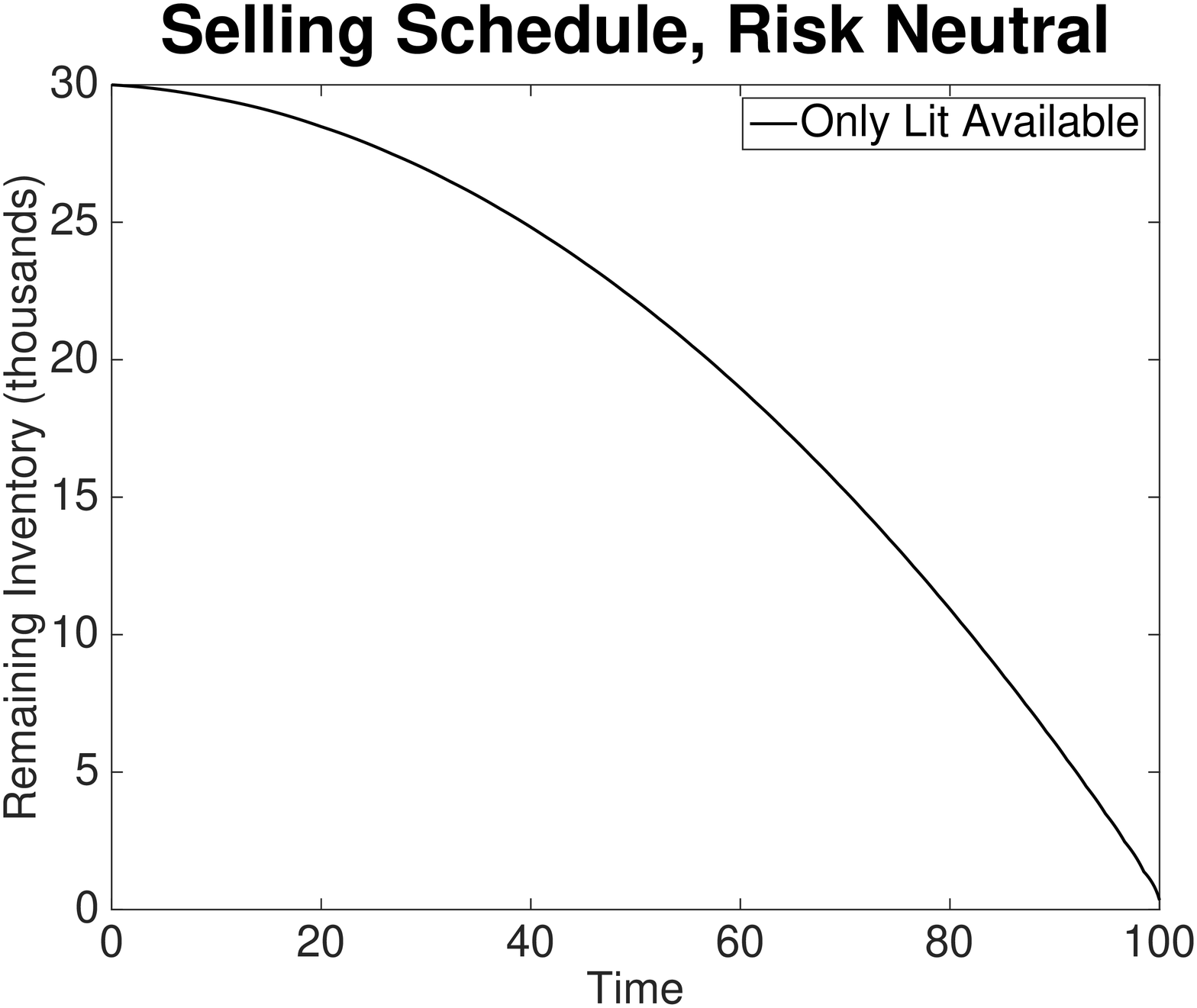}
\includegraphics[width=205pt,height=170pt]{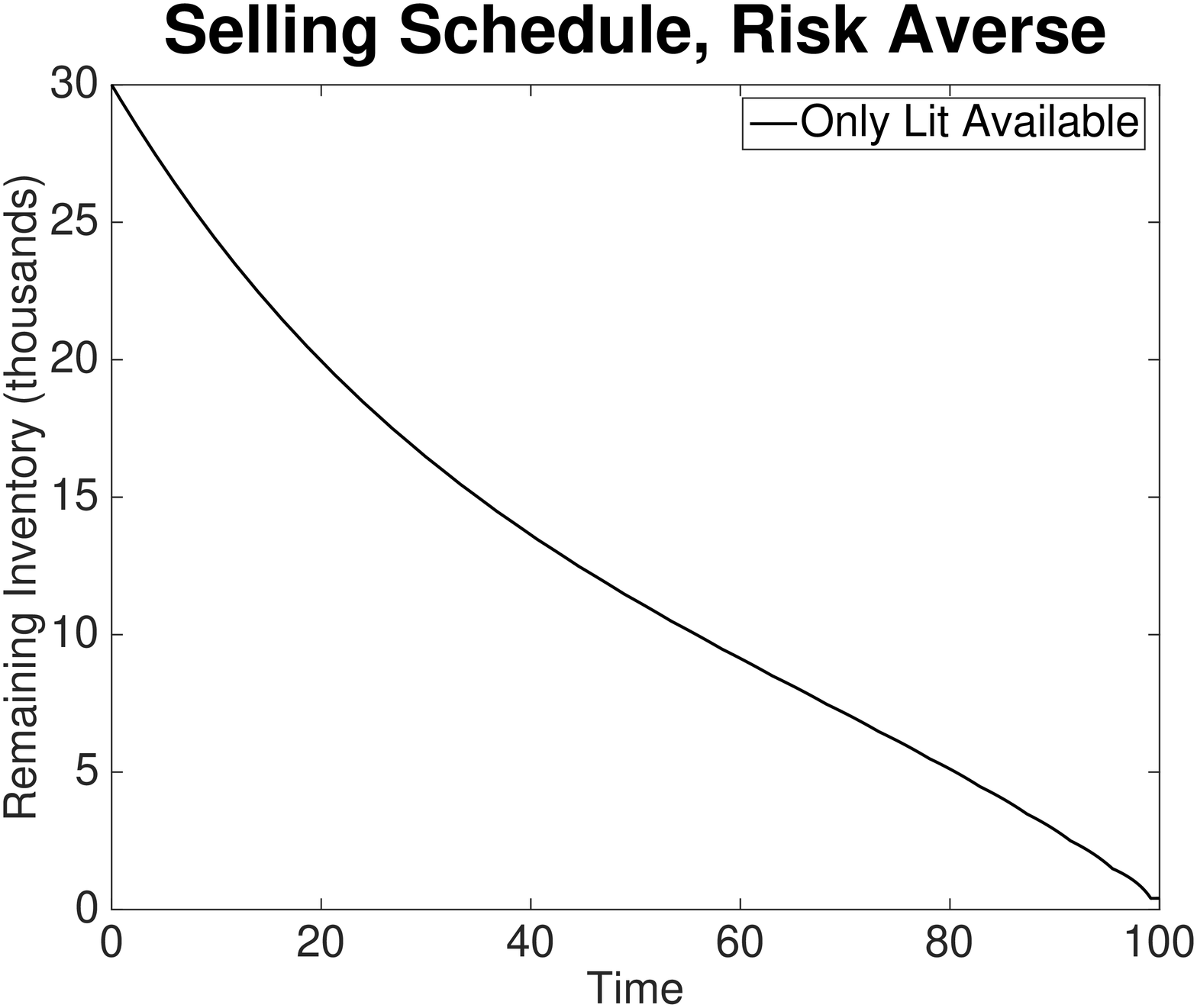}
\caption{\footnotesize{Optimal  strategy when the LOB best bid price is a submartingale. We set $\gamma=0$ (risk-neutral investor, left panel) , $\gamma=0.01$ (risk-averse investor, right panel). 
}}
\label{eq:lit1}
\end{figure}

\begin{figure}[H]
\centering
\includegraphics[width=205pt,height=170pt]{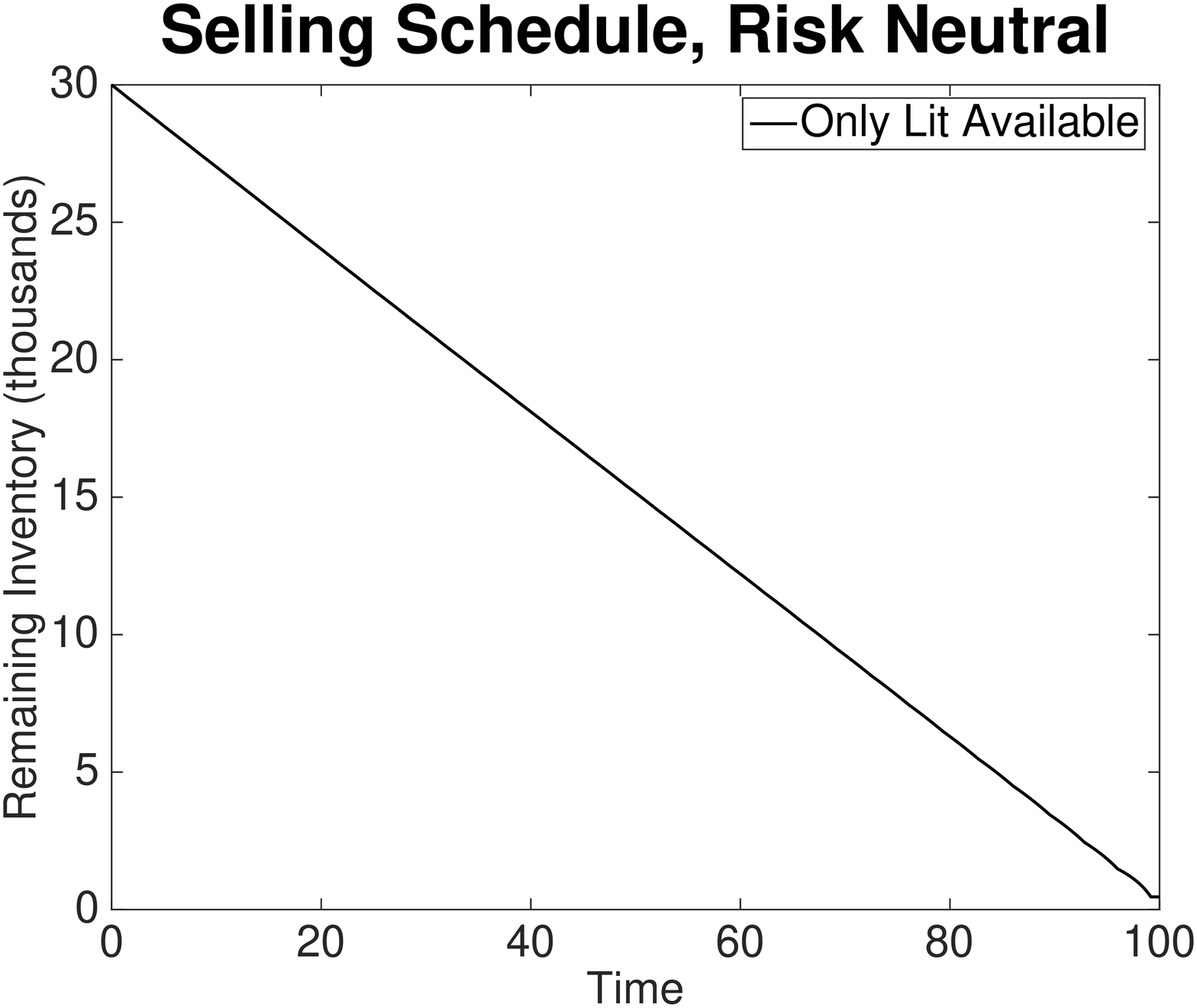}
\includegraphics[width=205pt,height=170pt]{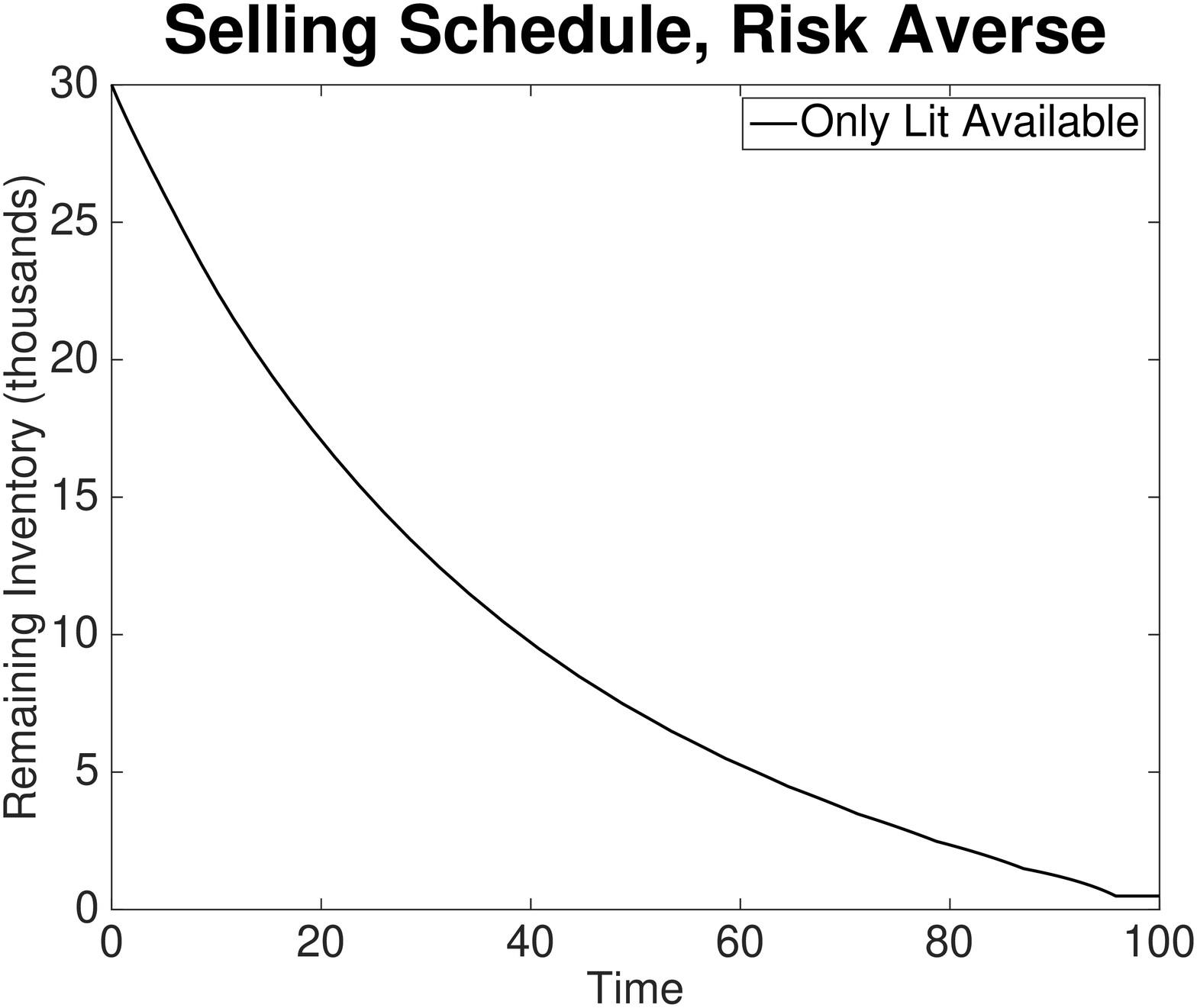}
\caption{\footnotesize{Optimal  strategy when the LOB best bid price is a martingale. We set $\gamma=0$ (risk-neutral investor, left panel) , $\gamma=0.01$ (risk-averse investor, right panel). }}
\label{eq:lit2}
\end{figure}

\begin{figure}[H]
\centering
\includegraphics[width=205pt,height=170pt]{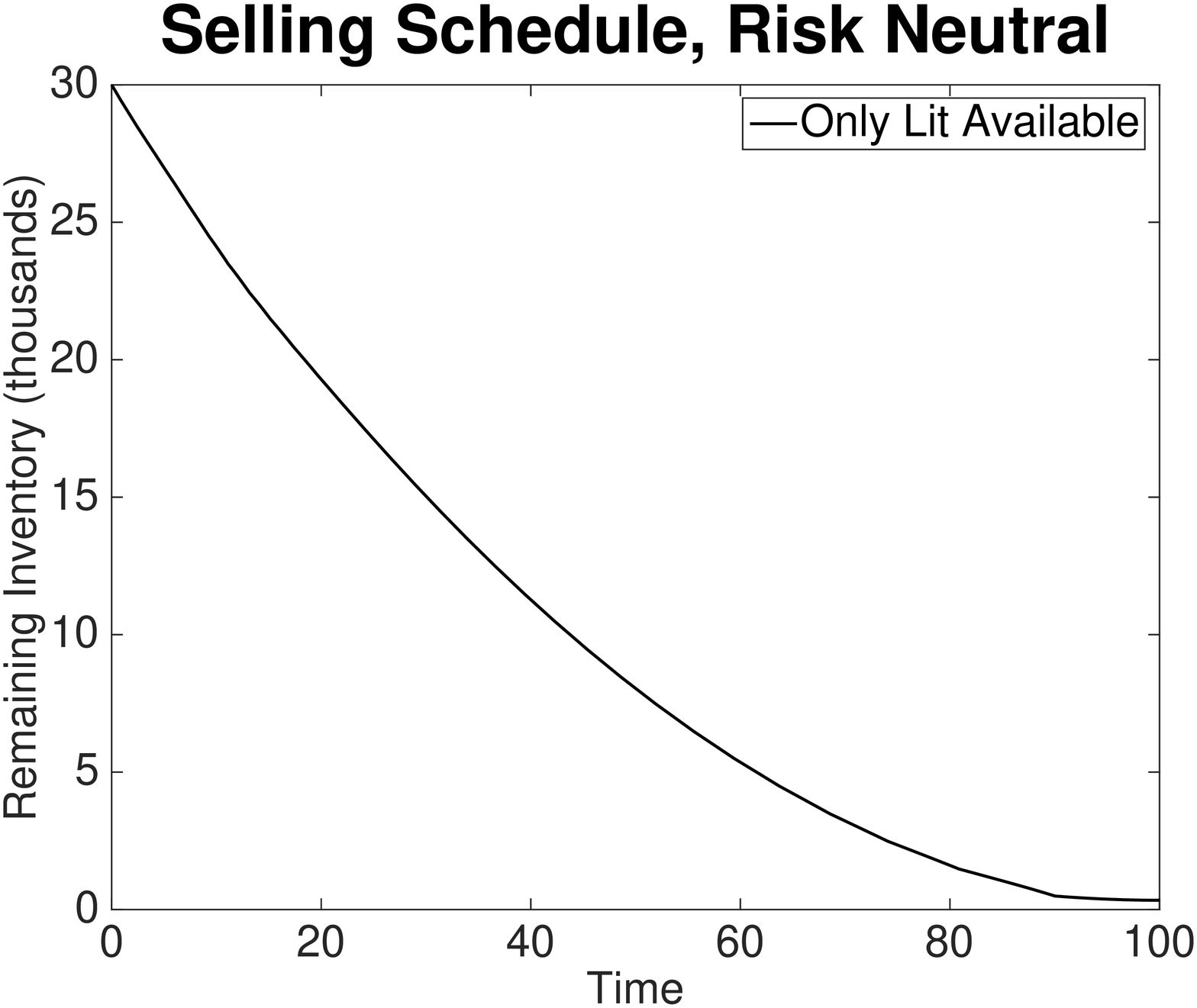}
\includegraphics[width=205pt,height=170pt]{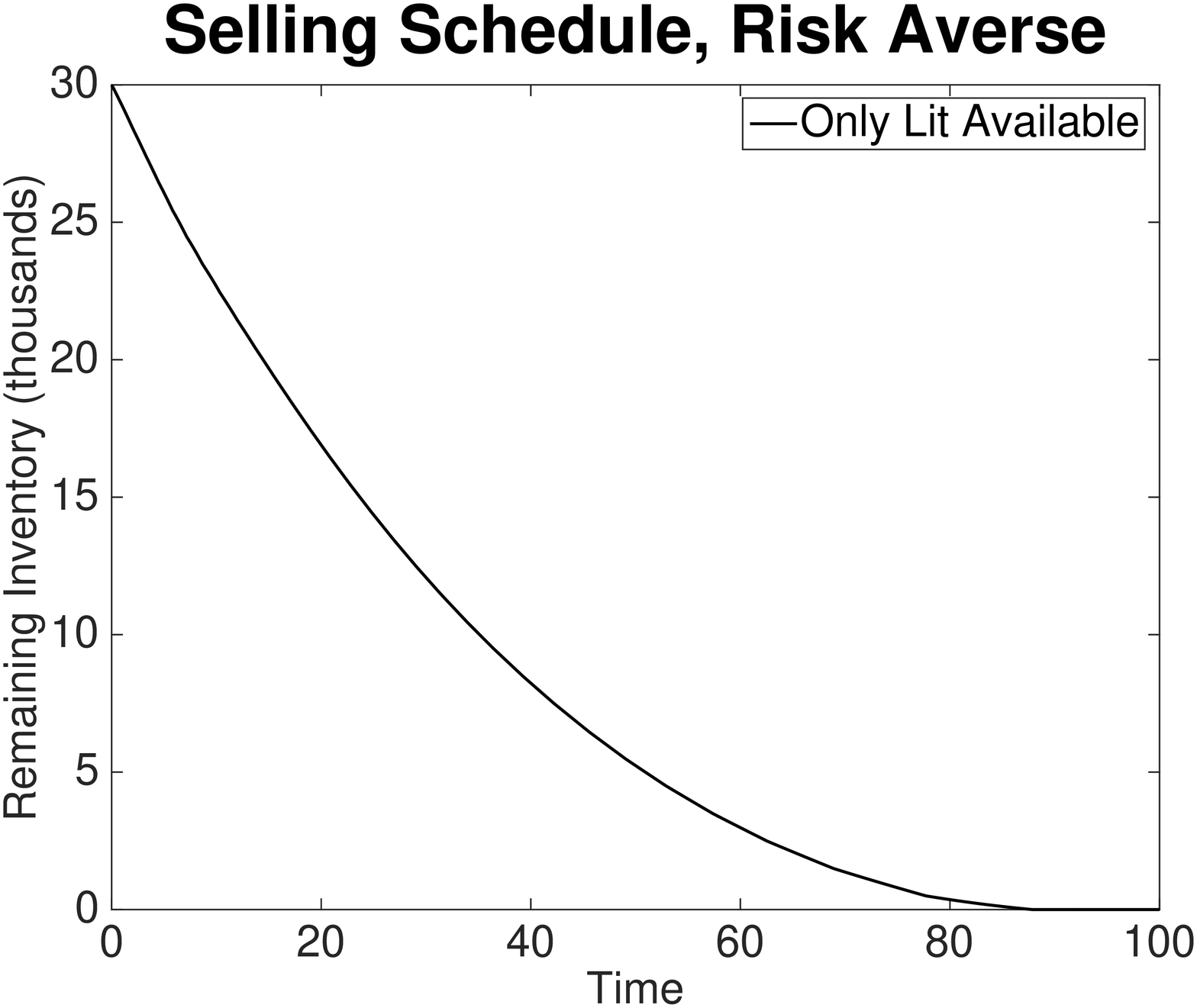}
\caption{\footnotesize{Optimal  strategy when the LOB best bid price is a supermartingale. We set $\gamma=0$ (risk-neutral investor, left panel) , $\gamma=0.01$ (risk-averse investor, right panel).}}
\label{eq:lit3}
\end{figure}

As well-known by practitioners and as also taken into account in much of the current literature (see e.g. Cartea et al. \cite{77}, and Fodra and Pham \cite{144}), the LOB mid-price mean-reverts quickly to its long-term mean. Thus we choose the following dynamics for the best bid price and the spread processes:

\begin{align}
\label{eq:bid1} \rd S^b\left(u\right)=&\ \kappa^{b}\left[\bar S-S^{b}(u)-\mu^b\nu\,(u)\right]  \rd u+\rd J_1^{b}(u)-\rd J_2^{b}(u),\\
 \label{eq:ask1} \rd \Delta\left(u\right)=&\ \-\kappa^{\Delta}\left[\bar \Delta-\Delta(u)+\mu^\Delta\nu\,(u)\right]  \rd u+\rd J_1^{\Delta}(u)-\rd J_2^{\Delta}(u)-\rd J_1^{b}(u)+\rd J_2^{b}(u), 
\end{align}
where, for $i=1,2$, $J^{b}_{i}(u)=\sum_{j=1}^{N^{b}_{i}(u)}z_{i,j}$ and $J^{\Delta}_{i}(u)=\sum_{j=1}^{N^{\Delta}_{i}(u)}y_{i,j}$ are compound Poisson processes with intensities $\lambda_{i}^{b}$ and $\lambda_{i}^{\Delta}$, respectively. For $j=1,2,\dots$, we let  $z_{i,j}$ and $y_{i,j}$ be sequences of non-negative i.i.d. random variables with bounded supports $Z_{i}$ and $Y_{i}$, respectively. In (\ref{eq:bid1}) we interpret $\rd J_1^{b}$ as the change in the best bid price due to the submission of limit buy orders at a more favourable price, whereas $\rd J_2^{b}$ models the changes due to incoming market sell orders which walk the book and the effect of cancellations of limit sell orders posted at the best price. The quantities $\mu^b$ and $\mu^\Delta$ model the permanent impact to which both, the best bid price and the spread in the standard exchange are subject to, after a trade takes place. In  Figure \ref{eq:gr1}, we plot a simulation of the best ask, the mid, and the best bid prices given respectively by Equations (\ref{eq:SaSm}),  (\ref{eq:middp}), and (\ref{eq:bid1}). 
\vspace{-0.38cm}
\noindent An assumption that is usually made, see e.g. Almgren and Chriss \cite{2}, is that the investor will get the trade orders executed at a price that includes an instantaneous impact commensurate to the liquidation rate $\nu\,(u)$. This feedback effect is commonly referred to as ``temporary impact''. The temporarily impacted best bid price $\hat{S}^b(u)$ and the wealth process are given by
$$
\hat{S}^b(u)=S^b(u)-\beta \nu\,(u),
$$ 
\begin{equation}
\rd W(u)=\nu\,(u)\hat S^{b}(u)\rd u+\eta(u)S^{m}(u^{-})\rd J(u),
\label{eq:wealth}
\end{equation}
where $S^{m}(u)$ is defined by Equation (\ref{eq:middp}). Next we consider an investor who wants to optimally liquidate his portfolio by placing sell orders in both, the standard market and the dark pool. The inventory process of the investor is here modelled by (\ref{inventdp}). The maximised expected return derived by the shares sale is obtained by solving the optimisation problem
\begin{equation}
\begin{split}
V\left(t,\boldsymbol{x}\right)=\sup_{\boldsymbol{\nu}\in\mathcal Z}\,&\E\left[W(T)+({S}^b(T)-\alpha X(T))X(T)-\gamma\int_{t}^\tau X^2(u)\rd u\right],
\label{eq:probl}
\end{split}
\end{equation}
where $\tau$ is defined by (\ref{eq:tau}) and $\gamma\in\RR_{+}\cup0$.  In the considered performance criterion, we allow for a maximisation of the terminal cash $W(T)$ together with the terminal value of the portfolio $S^{b}(T)X(T)$ and a penalty for a non-zero inventory level at time $T$ given by  $-\alpha X^{2}(T)$, where $\alpha>0$. The integral term, as in Cartea et al.  \cite{66,7},  penalises for the inventory holding over the whole period in which the strategy is applied. It is shown in the aforementioned papers that such a term can also be related to both, the variance of the portfolio value and the ambiguity aversion to the mid-price-drift. (In such a case, $\gamma$ represents the risk-aversion parameter.) The associated   HJB PIDE is given by 
\begin{equation}\label{MR-HJB}
\begin{split}
\sup_{\boldsymbol{v}\in\mathcal Z}\,\biggl\{-\gamma x^2\!+\frac{\partial V}{\partial t}\!\left(t,\boldsymbol{x}\right)+\kappa^{b}\left[\bar S-s^{b}-\mu^b v\right]\frac{\partial V}{\partial s^{b}}\!\left(t,\boldsymbol{x}\right)+\kappa^{\Delta}\left[\bar \Delta-\Delta+\mu^\Delta v\right]\frac{\partial V}{\partial \Delta}\!\left(t,\boldsymbol{x}\right)&\\
-v\frac{\partial V}{\partial x}\!\left(t,\boldsymbol{x}\right)+v(s^{b}-\beta v)\frac{\partial V}{\partial w}\!\left(t,\boldsymbol{x}\right)&\\
+\lambda^{w}\E\left[V\left(t,x-n z^{w},s^{b},\Delta, w+n z^{w}\left(s^{b}+\frac{\Delta}{2}\right)   \right)\!-\!V\left(t,\boldsymbol{x}\right)\right]&\\
+\lambda_1^{b}\E\Big[V\!\left(t,x,s^{b}+z_{1},\Delta-z_1, w\right)\!-\!V\left(t,\boldsymbol{x}\right)\Big]&\\
+\lambda_2^{b}\E\Big[V\!\left(t,x,s^{b}-z_{2},\Delta+z_2, w\right)\!-\!V\left(t,\boldsymbol{x}\right)\Big]&\\
+\lambda_1^{\Delta}\E\Big[V\!\left(t,x,s^{b}\!,\Delta+y_{1}, w\right)\!-\!V\left(t,\boldsymbol{x}\right)\Big]&\\
+\lambda_2^{\Delta}\E\Big[V\!\left(t,x,s^{b}\!,\Delta-y_{2}, w\right)\!-\!V\left(t,\boldsymbol{x}\right)\Big] &\biggr\}=0,
\end{split}
\end{equation}
\noindent with terminal condition $V\big(T,x,s^b,\Delta,w\big)=w+({s}^b-\alpha x)x$. We first show the optimal trading strategy, which we obtain numerically by solving (\ref{MR-HJB}), when only a standard exchange is available to the investor. This is achieved by imposing $z^{w}_{i}\equiv 0$, so to consider a dark pool that never executes orders (the dark pool is not available to the investor, that is). The following figures show how the optimal strategy varies depending on whether the mid-price process is a sub-, super-, or (local) martingale.

\vspace{-0.5cm}
\vspace{-0.5cm}

\begin{figure}[H]
\centering
\includegraphics[width=205pt,height=150pt]{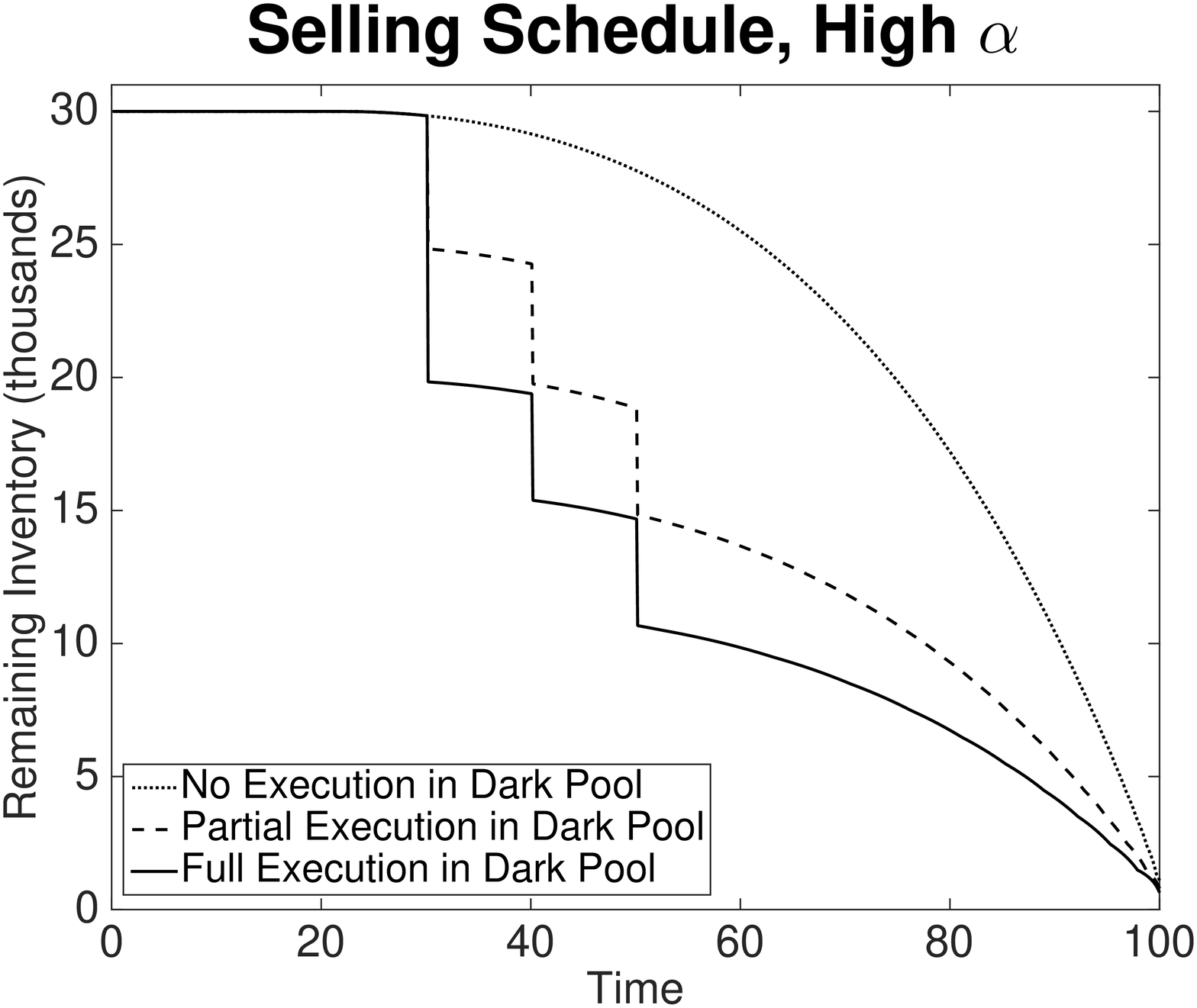}
\includegraphics[width=205pt,height=150pt]{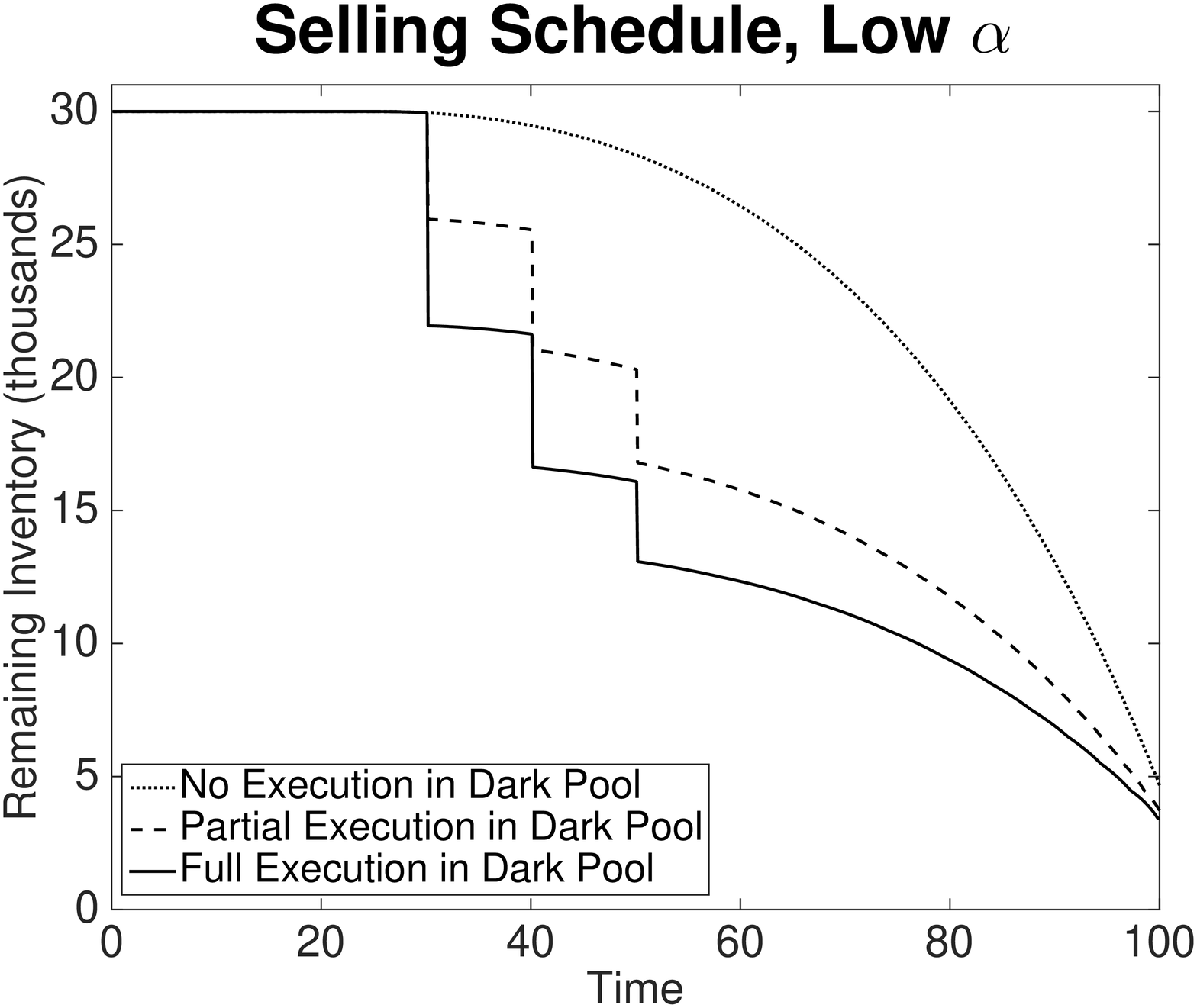}
\caption{\footnotesize{Optimal selling strategy displayed as a function of the remaining inventory. We set $\gamma=0.0001$, $\lambda_1^b=\lambda_2^b=\lambda^{\Delta}_1=\lambda^{\Delta}_2=0.2$, $z_{i},w_{i}\sim U[0,0.1]$, $s^{b}=40$, $\Delta=0.1$, $\bar \Delta=0.1$, $\bar S=40$, $\kappa^{b}=\kappa^\Delta=0.02$, $\beta=1\times10^{-5}$, $z^{w}\sim U[0,1]$, $\lambda^{w}=0.1$, $\mu^\Delta=\mu^b=0.01$. Left panel: $\alpha=6$. Right panel: $\alpha=0.5$.
}}

\label{eq:g222}
\end{figure}

\begin{figure}[H]
\centering
\includegraphics[width=205pt,height=150pt]{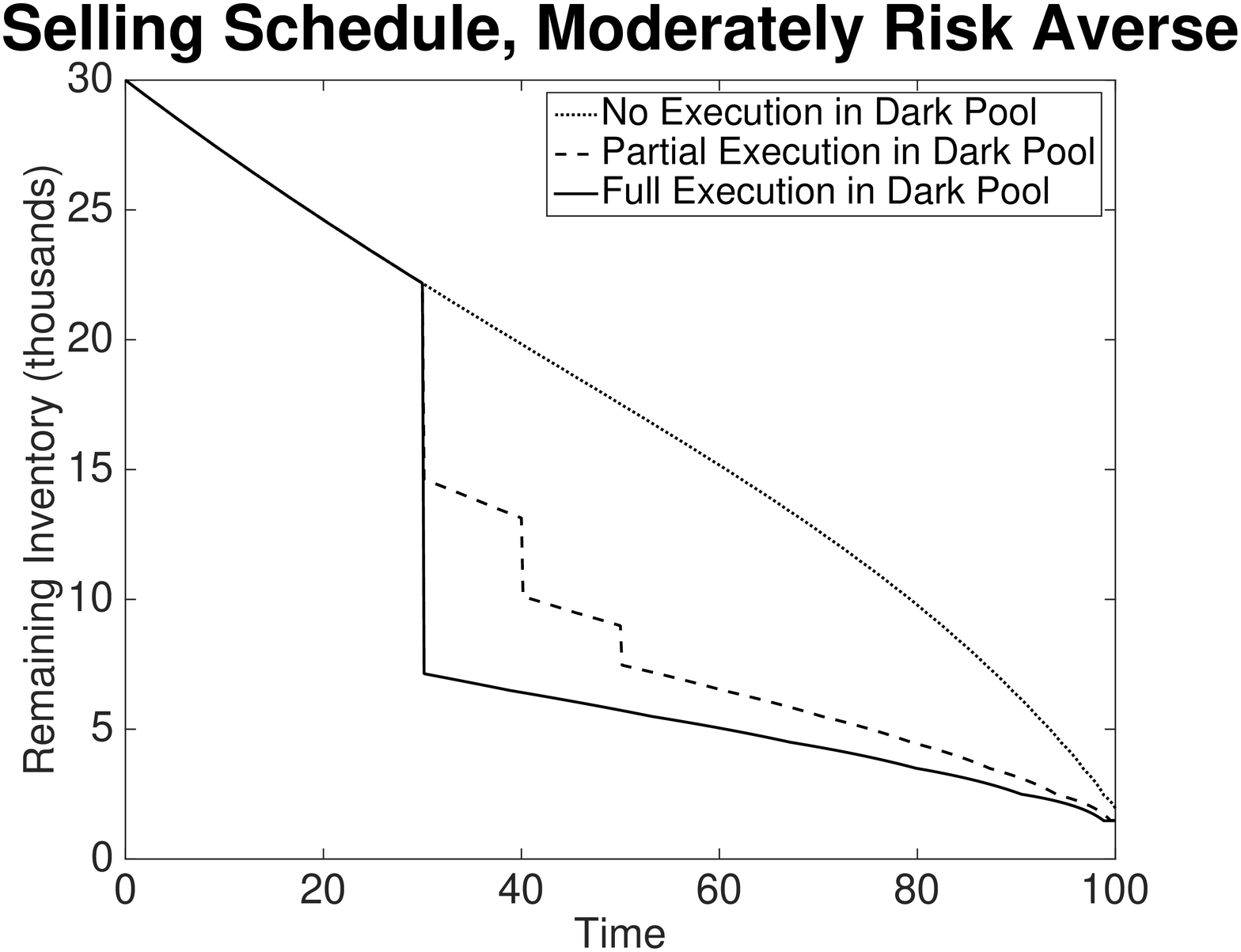}
\includegraphics[width=205pt,height=150pt]{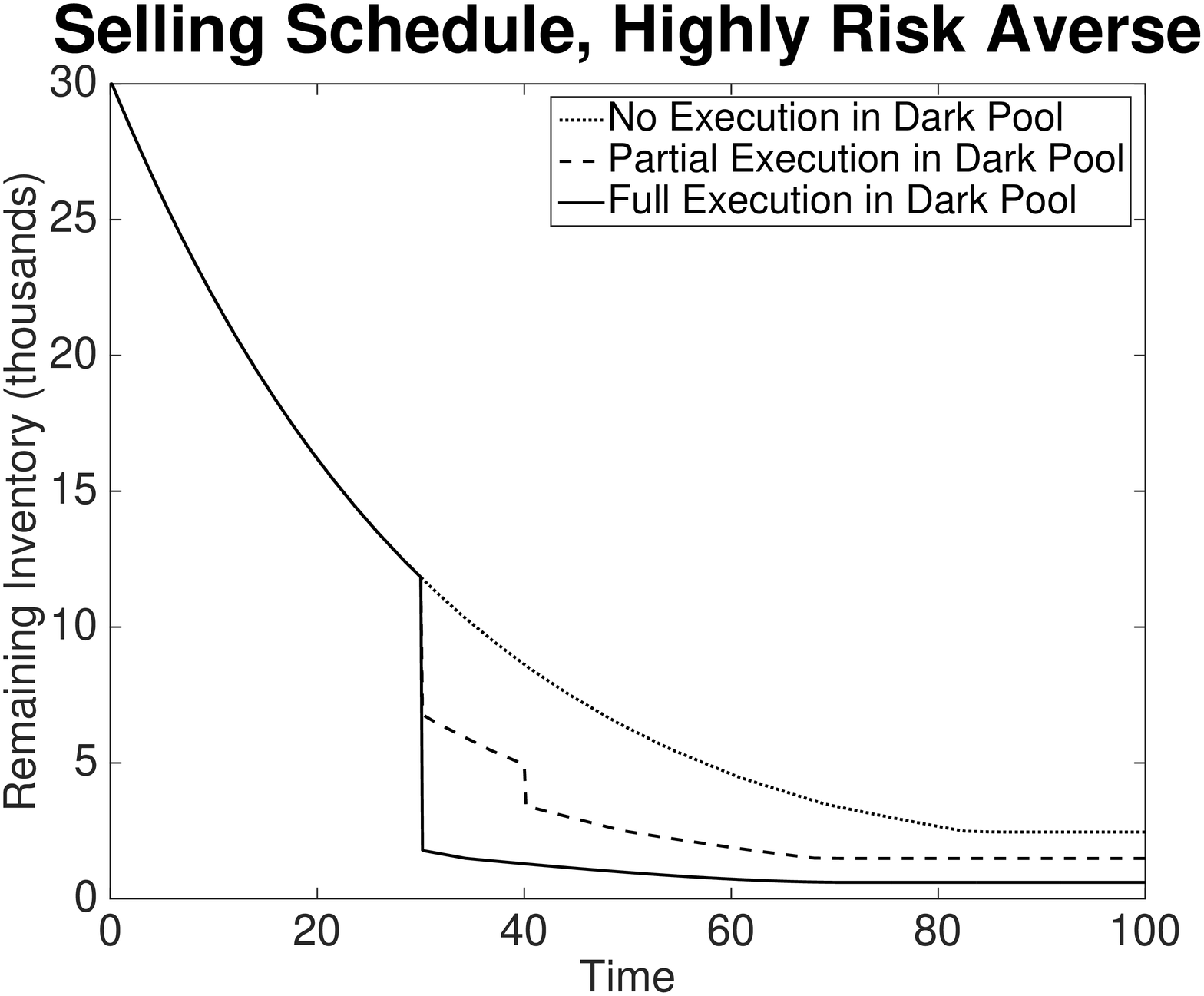}
\caption{\footnotesize{Optimal selling strategy displayed as a function of the remaining inventory. We set $\alpha=2$, $\lambda_1^b=\lambda_2^b=\lambda^{\Delta}_1=\lambda^{\Delta}_2=0.2$, $z_{i},w_{i}\sim U[0,0.1]$, $s^{b}=40$, $\Delta=0.1$, $\bar \Delta=0.1$, $\bar S=40$, $\kappa^{b}=\kappa^\Delta=0.02$, $\beta=1\times10^{-5}$, $z^{w}\sim U[0,1]$, $\lambda^{w}=0.1$, $\mu^\Delta=\mu^b=0.01$. Left panel: $\gamma=0.01$. Right panel: $\gamma=0.1$.
}}
\label{eq:g333}
\end{figure}

\begin{figure}[H]
\centering
\includegraphics[width=205pt,height=150pt]{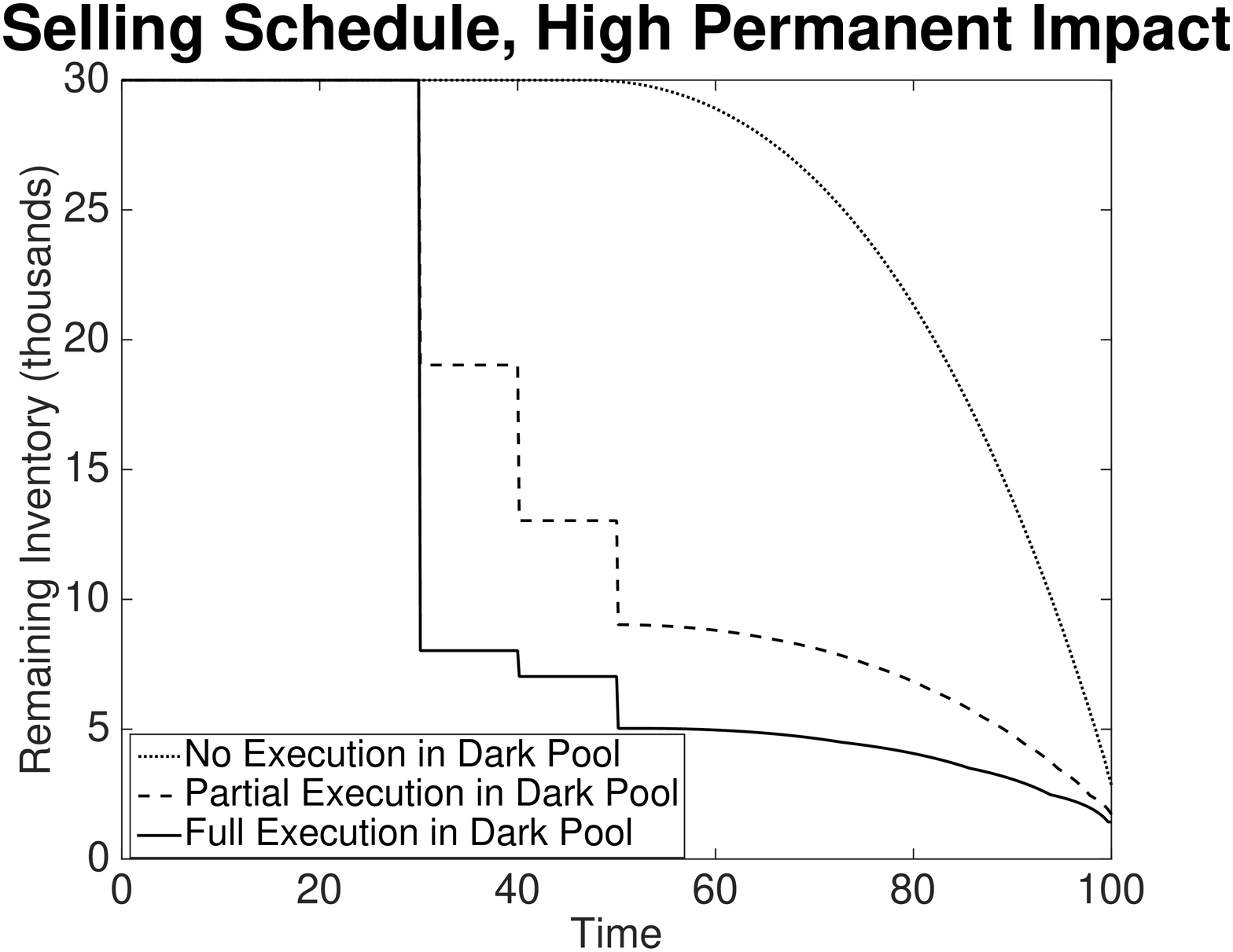}
\includegraphics[width=205pt,height=150pt]{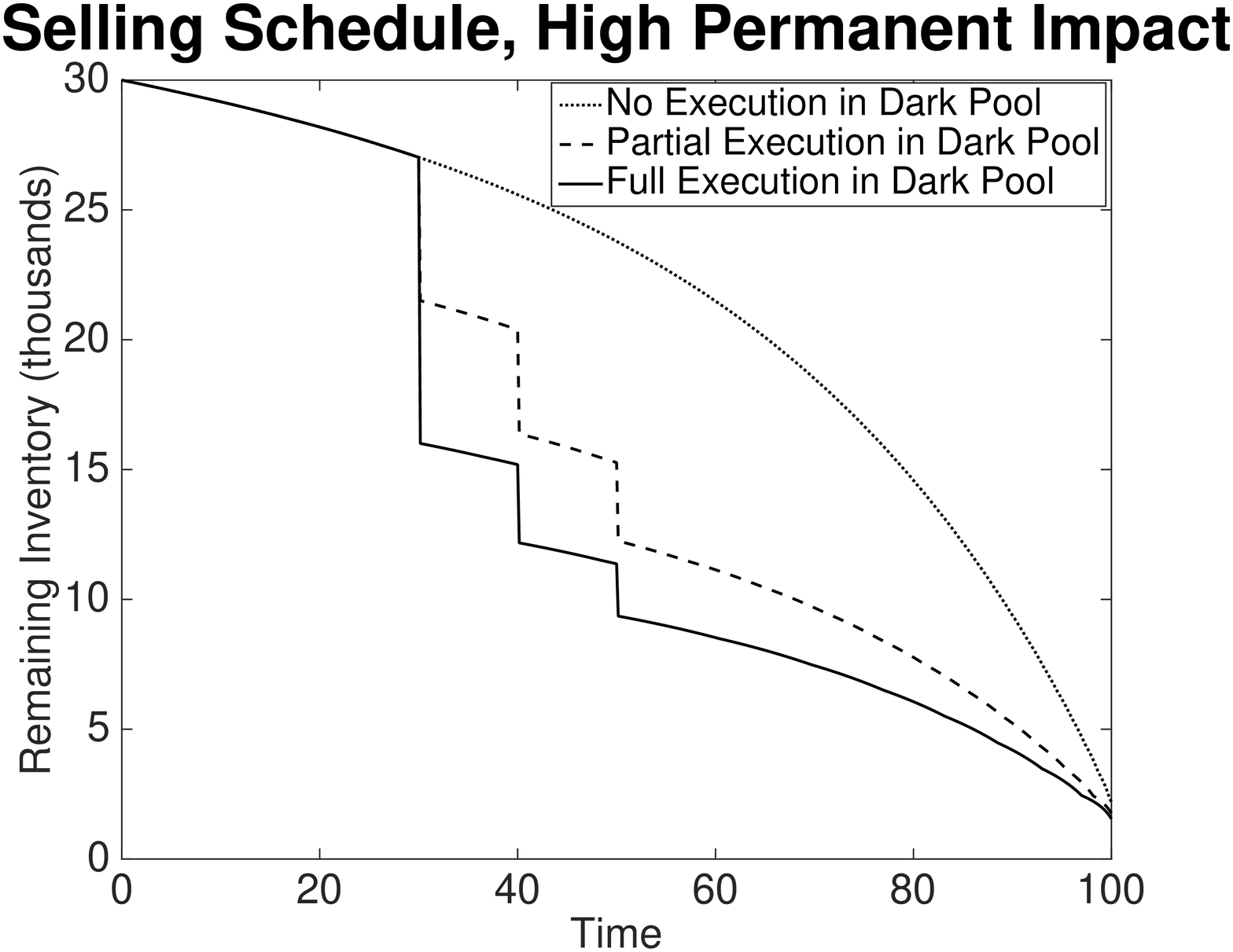}
\caption{\footnotesize{Optimal selling strategy displayed as a function of the remaining inventory. We set $\alpha=2$, $\gamma=0.0001$, $\lambda_1^b=\lambda_2^b=\lambda^{\Delta}_1=\lambda^{\Delta}_2=0.2$, $z_{i},w_{i}\sim U[0,0.1]$, $s^{b}=40$, $\Delta=0.1$, $\bar \Delta=0.1$, $\bar S=40$, $\kappa^{b}=\kappa^\Delta=0.02$, $\beta=1\times10^{-5}$, $z^{w}\sim U[0,1]$, $\lambda^{w}=0.1$. Left panel: $\mu^\Delta=\mu^b=0.01$. Right panel: $\mu^\Delta=\mu^b=0.0001$.
}}
\label{eq:g4444}
\end{figure}

In Figure \ref{eq:g222}, we  show the evolution of the inventory, throughout the liquidation period, for a risk-neutral investor (i.e. $\gamma=0$). We obtain the optimal trading strategy by solving the HJB PIDE numerically and we plot the case of multiple executions (occurring at $\tau_{1}=30$, $\tau_{2}=40$ and $\tau_{3}=50$, right panel). We emphasise that  $\tau_{1}$, $\tau_{2}$ and $\tau_{3}$ are fixed arbitrarily---for the sake of illustration only---after a complete solution has been found. The dotted line shows the evolution of the inventory when there is no dark-pool execution over the entire period (although the dark pool is available to the investor). The dashed line shows the results for partial execution in the dark pool, while the solid line shows the results when the posted order in the dark pool is fully executed. We first look at the role played by $\alpha$.
The parameter $\alpha$ models the terminal penalty for failing to liquidate the whole inventory by time $T$. Higher values of $\alpha$ incentivise the agent to increase the selling rate in the standard exchange and the size of the dark pool posting (left panel). On the contrary, smaller values of $\alpha$ allow the agent to retain a larger portion of inventory by terminal date $T$. The important feature of the resulting strategy is that it may be suboptimal to place all the remaining shares in the dark pool at every point in time. This is due to the LOB dynamics specified above. In Figure \ref{eq:g333}, we show how the strategy changes for a risk-averse investor ($\gamma>0$). A higher risk aversion (right panel) incentivises the agent to liquidate faster as he is more sensitive to potential movements of the market price, while a lower degree of risk aversion is reflected in a slower liquidation (left panel).
In Figure \ref{eq:g4444} we show the sensitivity of the strategy with respect to the permanent price impact. Depending on the market conditions, it may be worth waiting before posting orders in the standard exchange (when there is a high permanent impact) and hope to be executed in the dark pool, while accelerating the lit pool trading is optimal for the case of a more liquid market (less permanent price impact).

\begin{figure}[H]
\centering
\includegraphics[width=205pt,height=150pt]{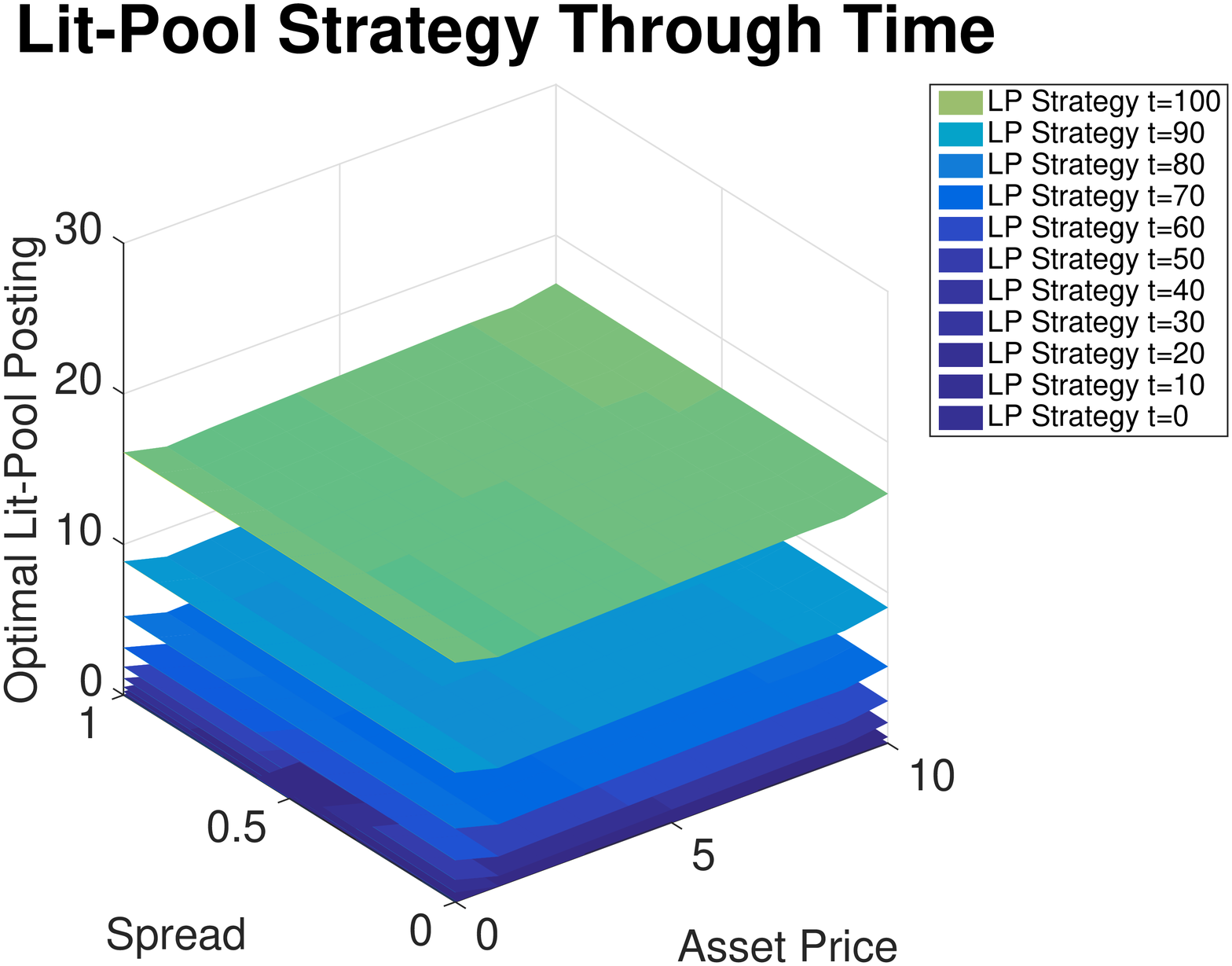}
\includegraphics[width=205pt,height=150pt]{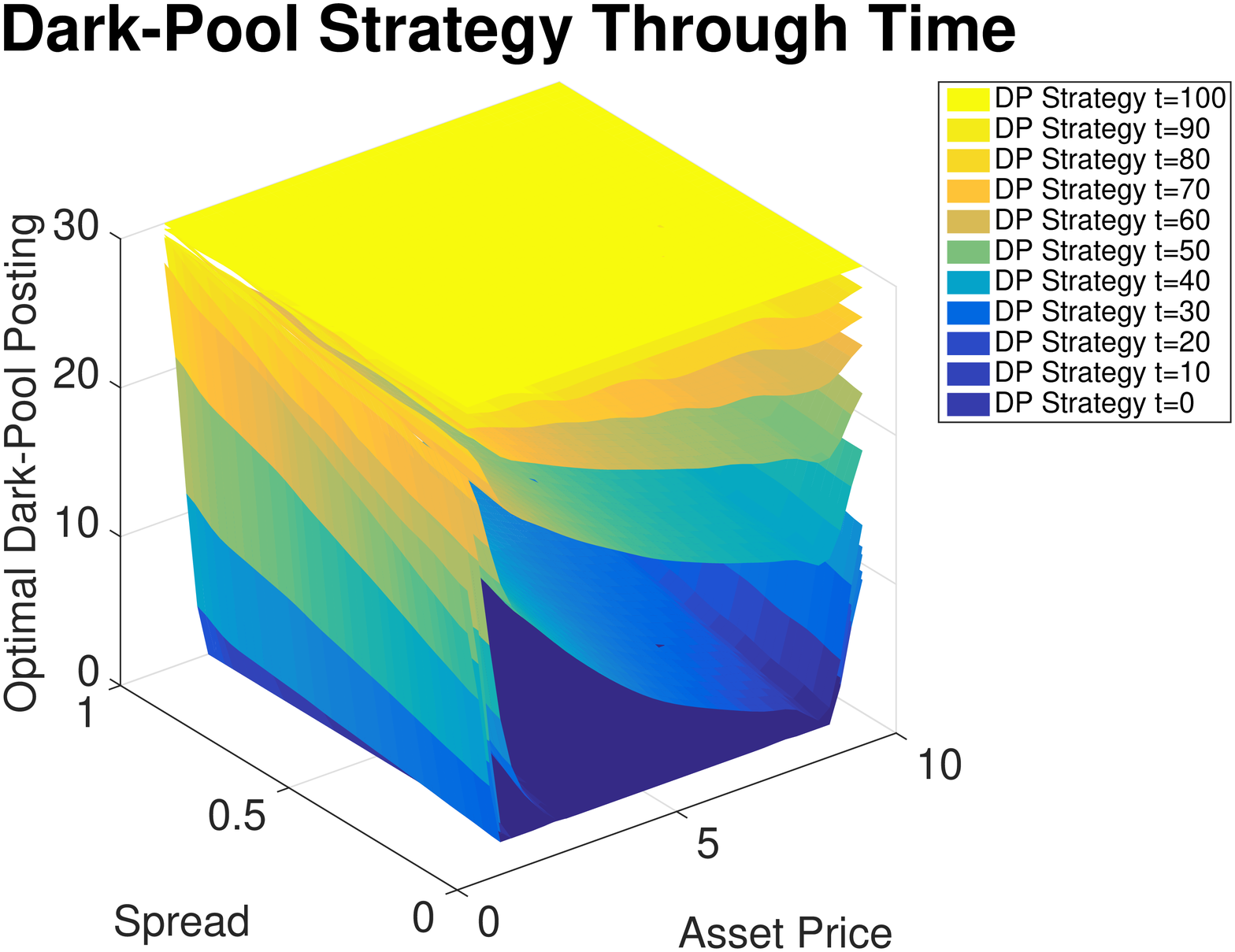}
\caption{\footnotesize{Optimal lit and dark pool strategies. We set $X_t=30,000$ for each $t$, $\alpha=2$, $\gamma=0.0001$, $\lambda_1^b=0.5$, $\lambda_2^b=0.1$, $\lambda^{\Delta}_1=\lambda^{\Delta}_2=0.2$, $z_{i},w_{i}\sim U[0,0.1]$, $\bar \Delta=0.1$, $\bar S=40$, $\kappa^{b}=\kappa^\Delta=0.02$, $\beta=1\times10^{-5}$, $z^{w}\sim U[0,1]$, $\lambda^{w}=0.1$, $\mu^\Delta=\mu^b=0.001$.
}}
\label{eq:round}
\end{figure}

\begin{figure}[H]
\centering
\includegraphics[width=205pt,height=140pt]{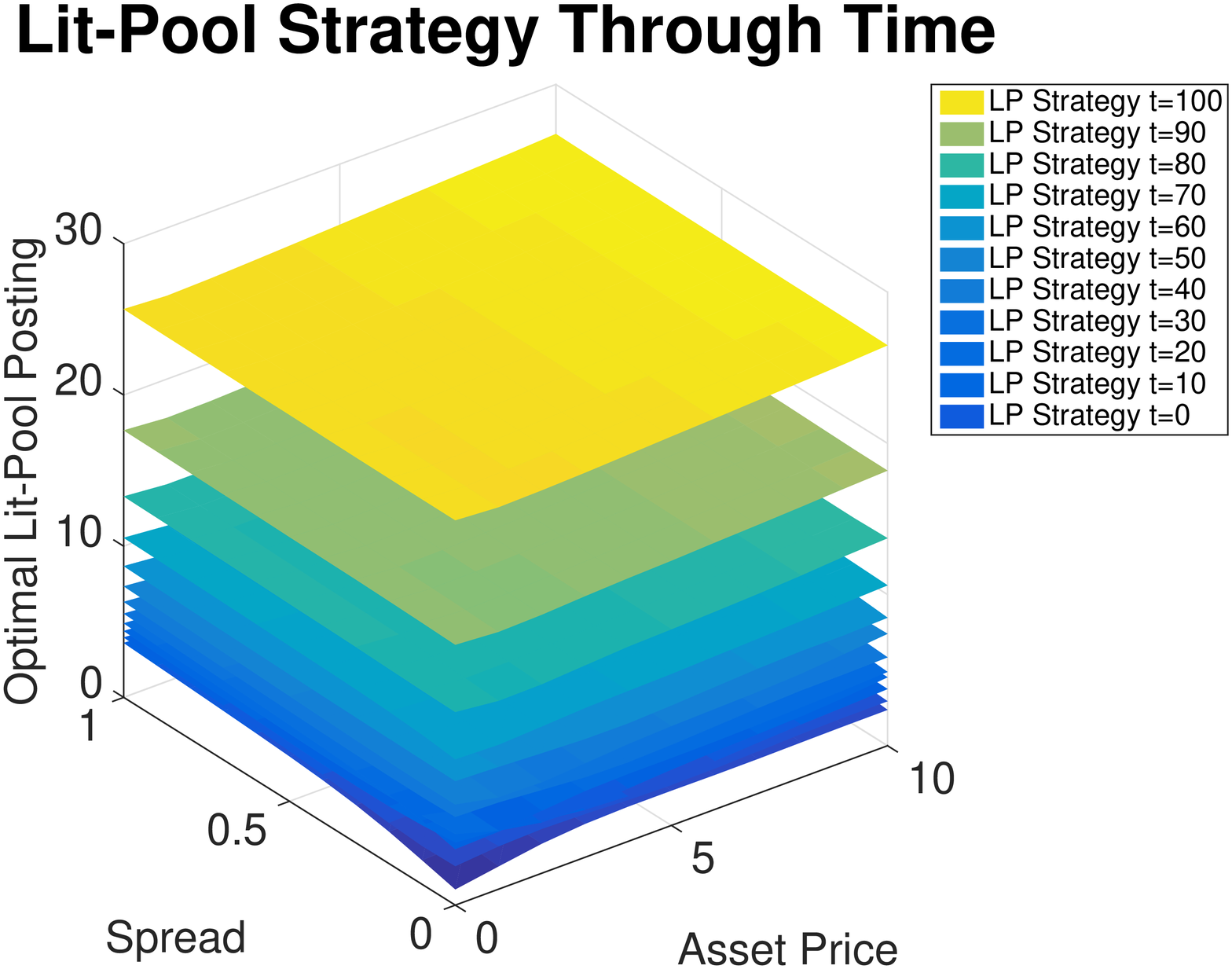}
\includegraphics[width=205pt,height=140pt]{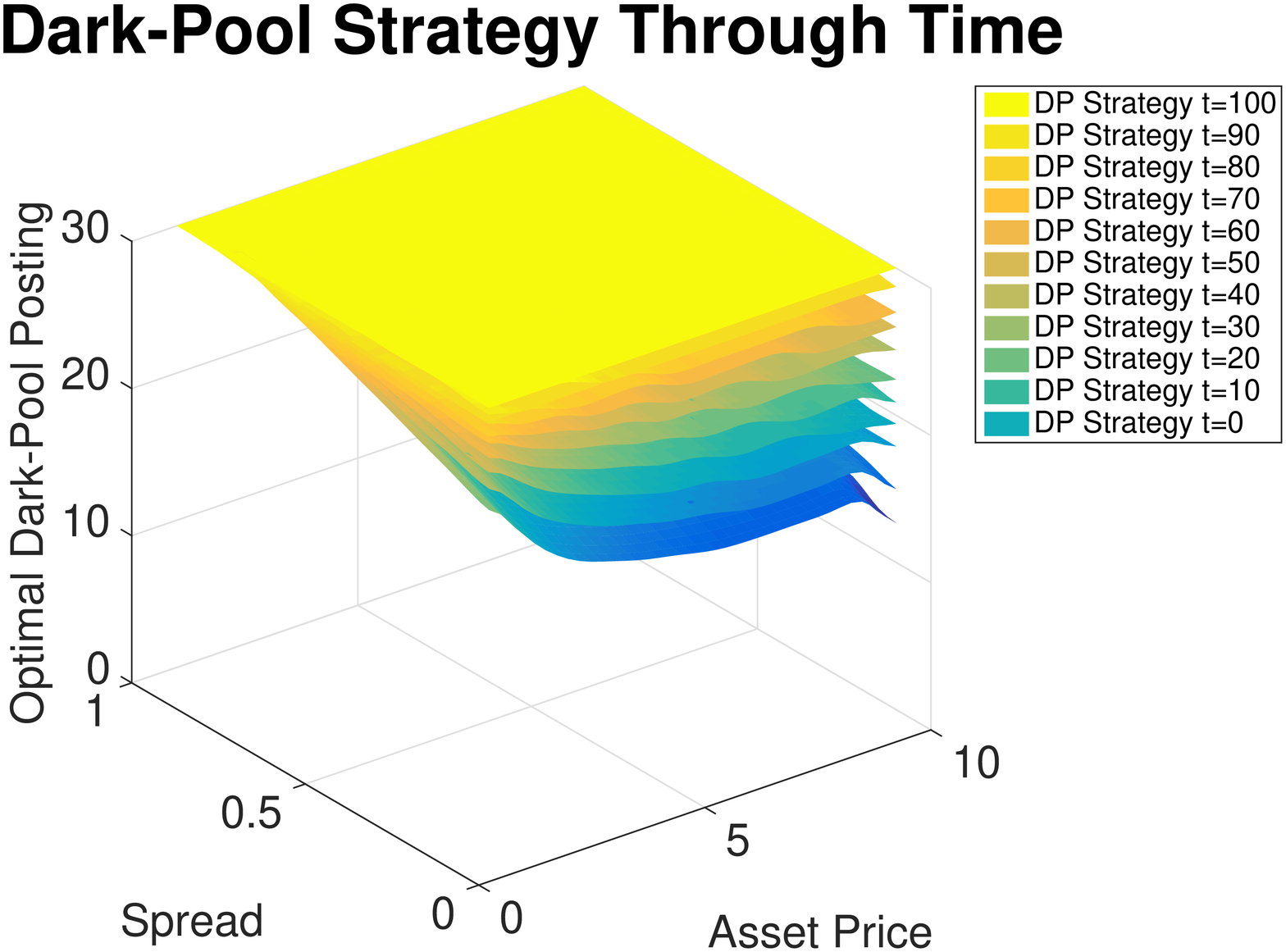}
\caption{\footnotesize{Optimal lit and dark pool strategies. We set $X_t=30,000$ for each $t$, $\alpha=2$, $\gamma=0.0001$, $\lambda_1^b=0.1$, $\lambda_2^b=0.5$, $\lambda^{\Delta}_1=\lambda^{\Delta}_2=0.2$, $z_{i},w_{i}\sim U[0,0.1]$, $\bar \Delta=0.1$, $\bar S=40$, $\kappa^{b}=\kappa^\Delta=0.02$, $\beta=1\times10^{-5}$, $z^{w}\sim U[0,1]$, $\lambda^{w}=0.1$, $\mu^\Delta=\mu^b=0.001$.
}}
\label{eq:round1}
\end{figure}

Roundtrips in the dark pool are not necessarily beneficial, that is, the agent may not wish to post all the remaining inventory in the dark pool. We have investigated such a feature by looking at fixed times $\tau$, but now we plot the whole optimal strategy $\nu$ and $\eta$ (i.e.) as a function of the bid price and the spread processes. This confirms our intuition that the agent may at times retain part of the inventory, depending on the market conditions. In Figure \ref{eq:round} we show the optimal strategy for both the lit and dark pools in the case that $\lambda_1^b>\lambda_2^b$, while in Figure \ref{eq:round1} we consider the case where $\lambda_1^b<\lambda_2^b$. When the price is expected to increase (Figure \ref{eq:round})  we notice that both strategies are reduced in size compared to Figure \ref{eq:round1}. This is coherent with the intuition that the agent may wish to exploit the opportunity of a price increase by waiting to liquidate the inventory.  We further see that the quantity posted in both venues increases as we approach the terminal date $T$. Take, e.g., the  left panel in Figure \ref{eq:round}. The bottom surface is the optimal trading rate in the standard exchange at the beginning of the trading period ($t=0$) with 30,000 shares to liquidate. As $t$ increases, the respective surface lifts up indicating that having less  time remaining implies posting higher quantities in the venue---we assume that the inventory $X=30,000$ at each time. The analogous holds for the right panel, in which we plot the dark pool strategy.

\subsection{Geometric L\'evy model}
\begin{figure}[H]
\begin{center}
\includegraphics[width=340pt,height=190pt]{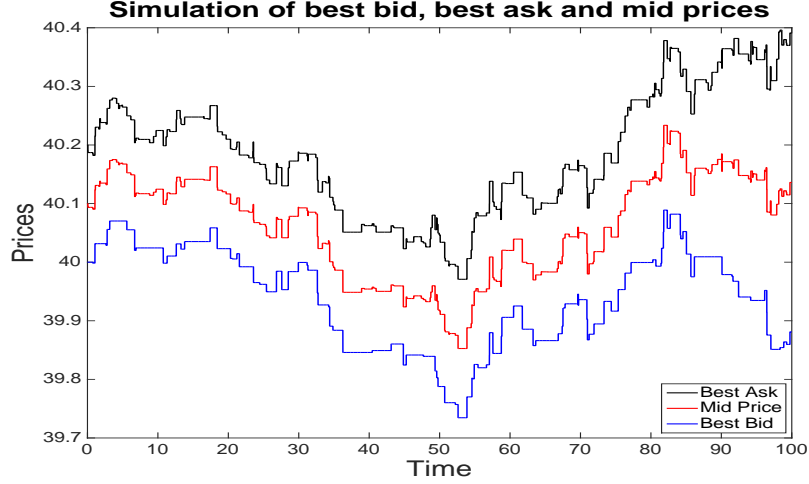}
\caption{\footnotesize{Simulation of the best ask, best bid and mid prices for a time frame of 100 seconds. We set $\lambda_1^b=\lambda_2^b=\lambda^{\Delta}_1=\lambda^{\Delta}_2=0.5$, $z_{i},y_{i}\sim U[0,0.1]$, $s^{b}=40$, $\Delta=0.2$.
}}
\label{eq:gr5}
\end{center}
\end{figure}

\begin{figure}[H]
\centering
\includegraphics[width=190pt,height=125pt]{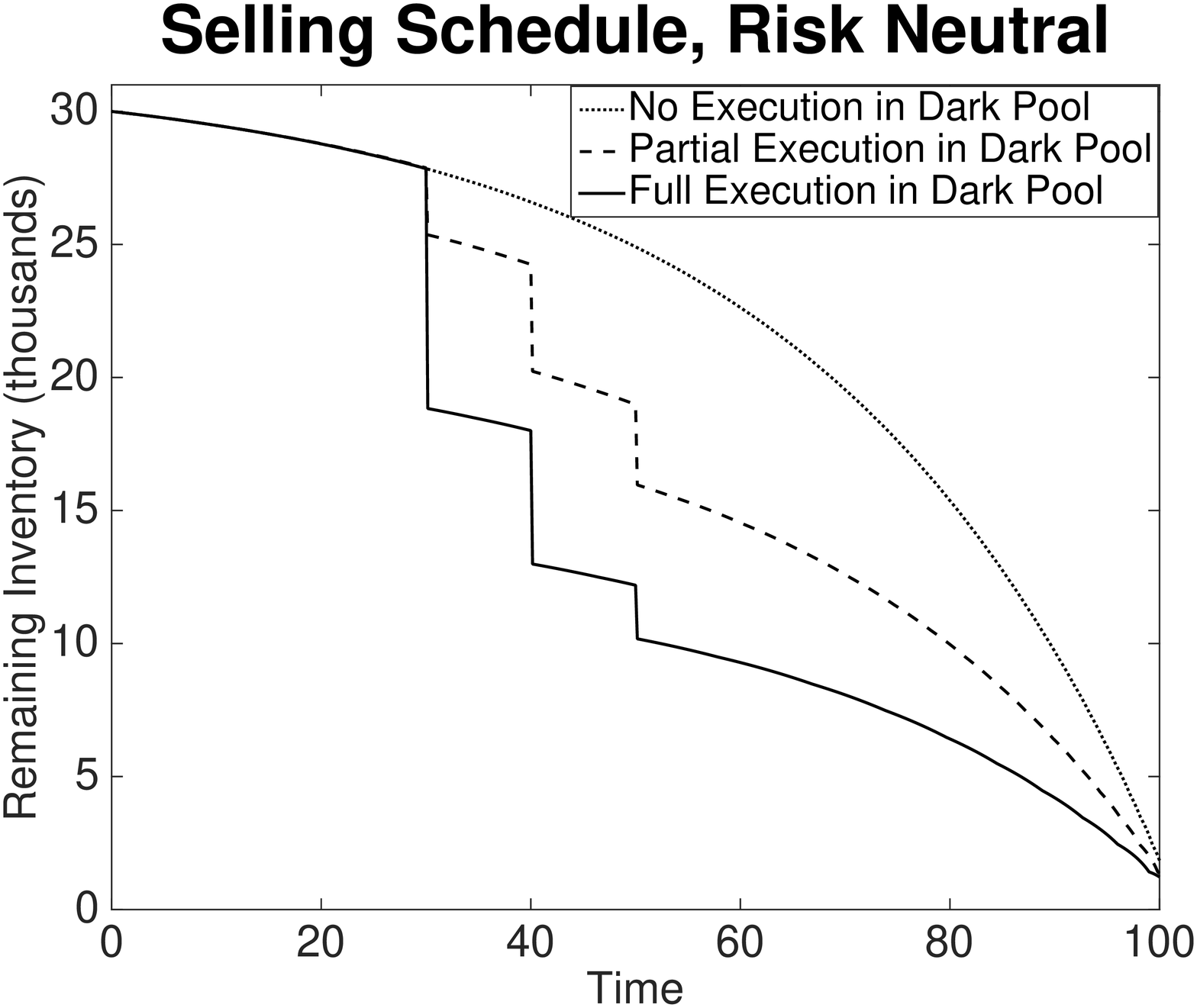}
\includegraphics[width=190pt,height=125pt]{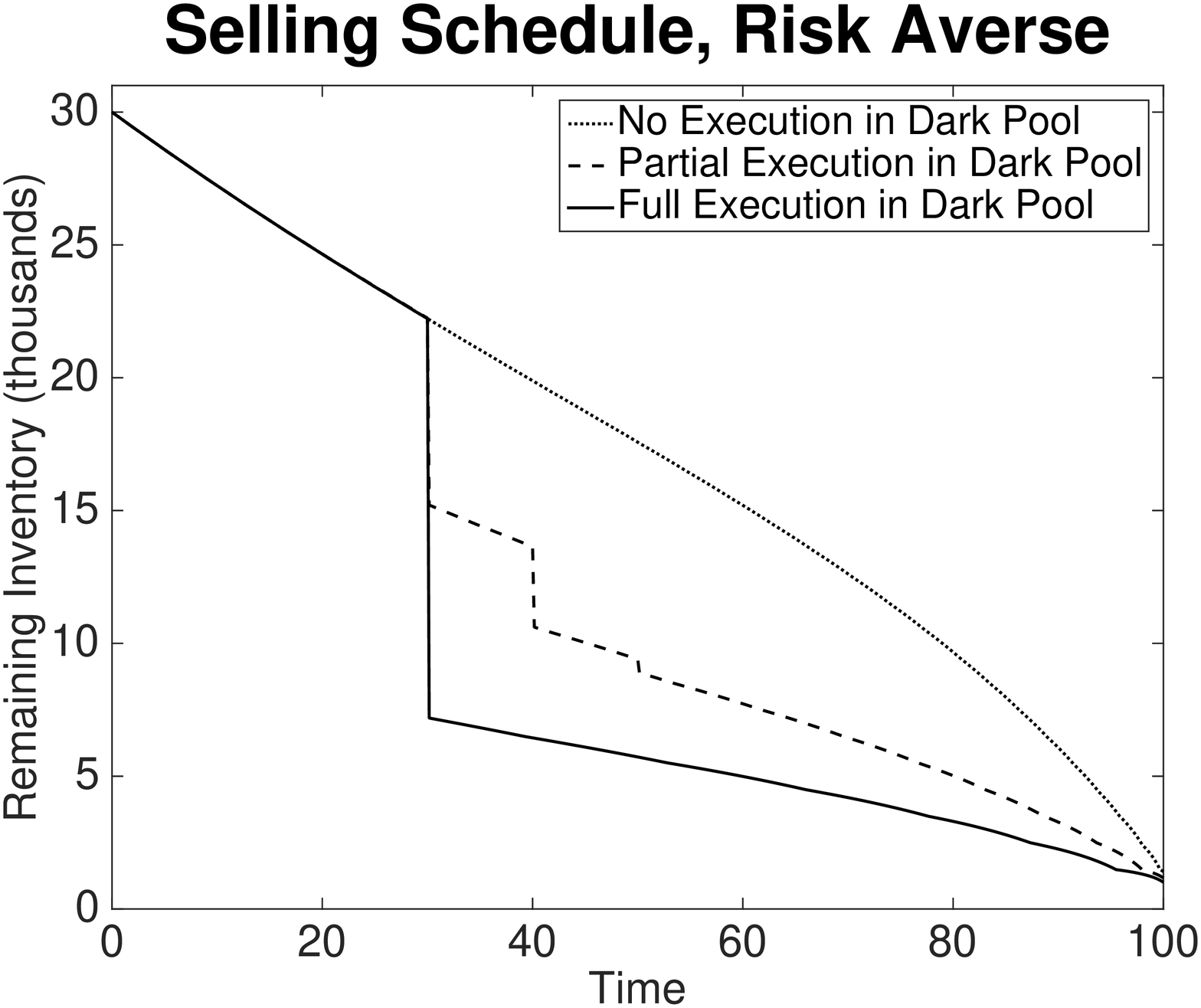}
\includegraphics[width=190pt,height=125pt]{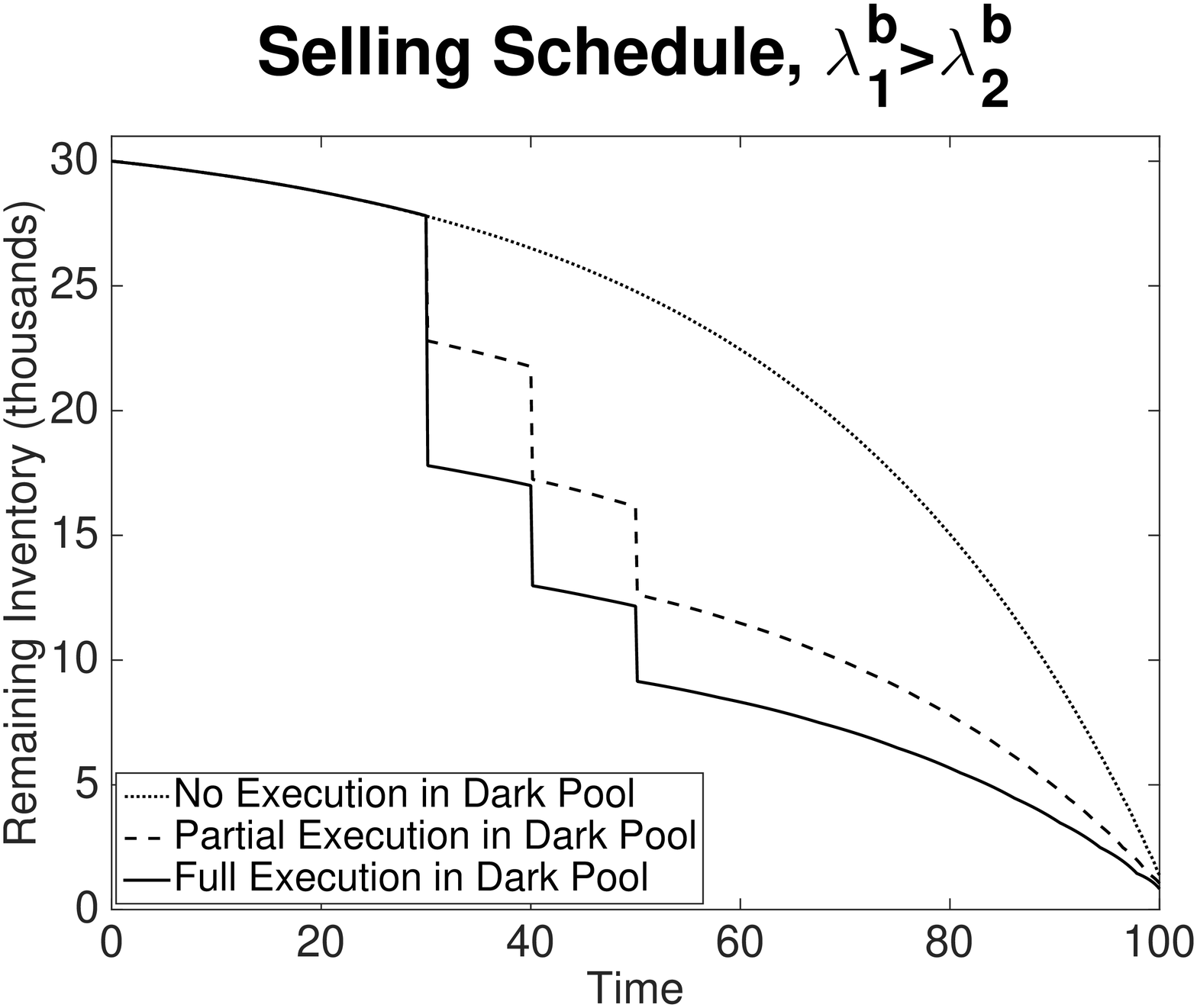}
\includegraphics[width=190pt,height=125pt]{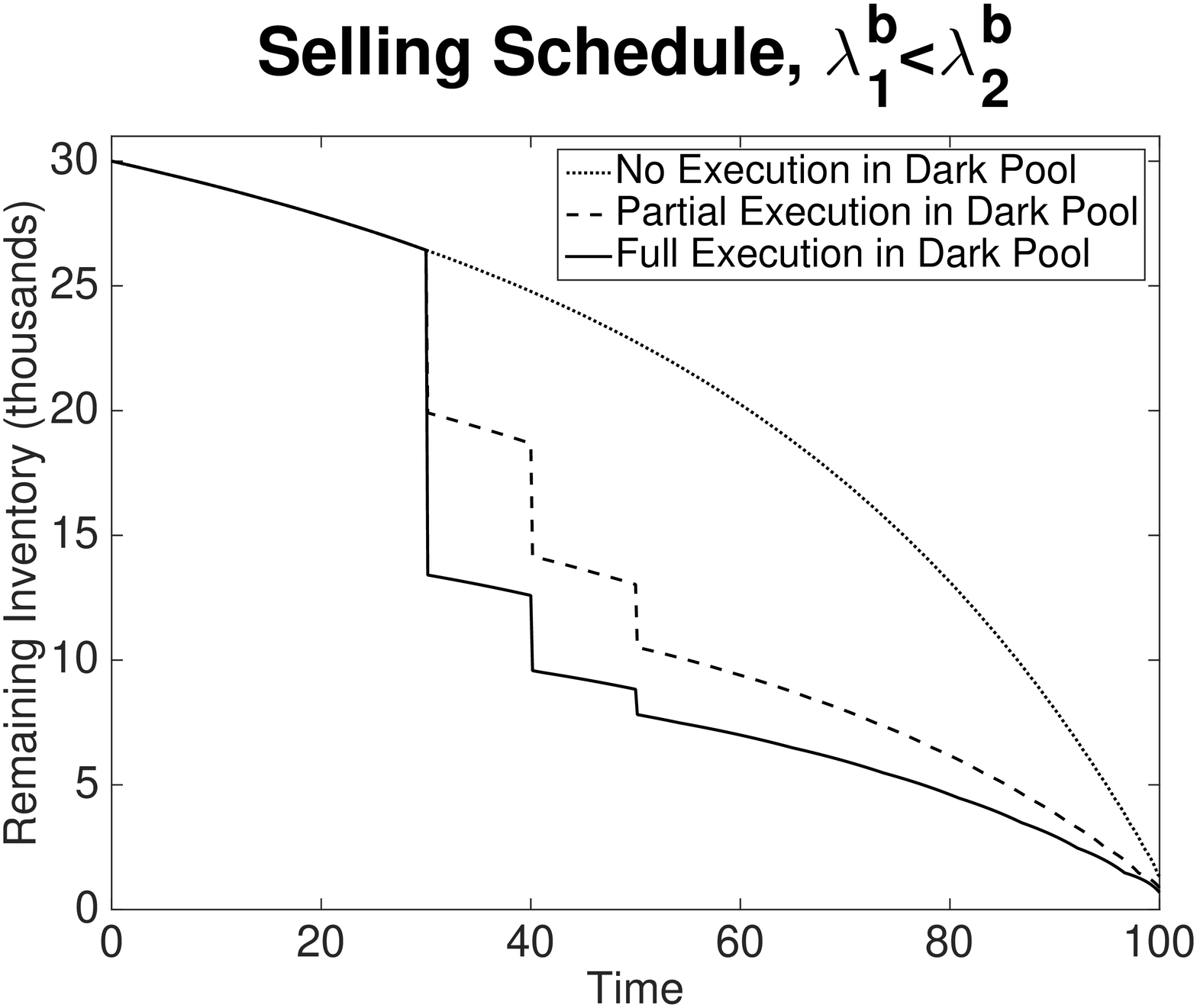}
\caption{{\footnotesize{Optimal selling strategy displayed as a function of the remaining inventory. We set $z_{i},y_{i}\sim U[0,0.1]$, $s^{b}=40$, $\Delta=0.3$, $\mu^{b}=0\mu^{\Delta}=0.0001$, $\beta=1\times10^{-5}$, $\alpha=2$, $z^{w}\sim U[0,1]$, $\lambda^{w}=0.1$. Top panels: $\lambda^b_1=\lambda^b_2=\lambda^\Delta_1=\lambda^\Delta_2=0.3$ and $\gamma=0$ (left), $\gamma=0.01$ (right). Bottom panels:  $\gamma=0.0001$ and $\lambda_1^b=\lambda_2^\Delta=0.5$, $\lambda^{b}_2=\lambda^{\Delta}_1=0.1$ (left), $\lambda_1^b=\lambda_2^\Delta=0.1$, $\lambda^{b}_2=\lambda^{\Delta}_1=0.5$.}}}
\label{eq:ag555}
\end{figure}

\begin{figure}[H]
\centering
\includegraphics[width=190pt,height=125pt]{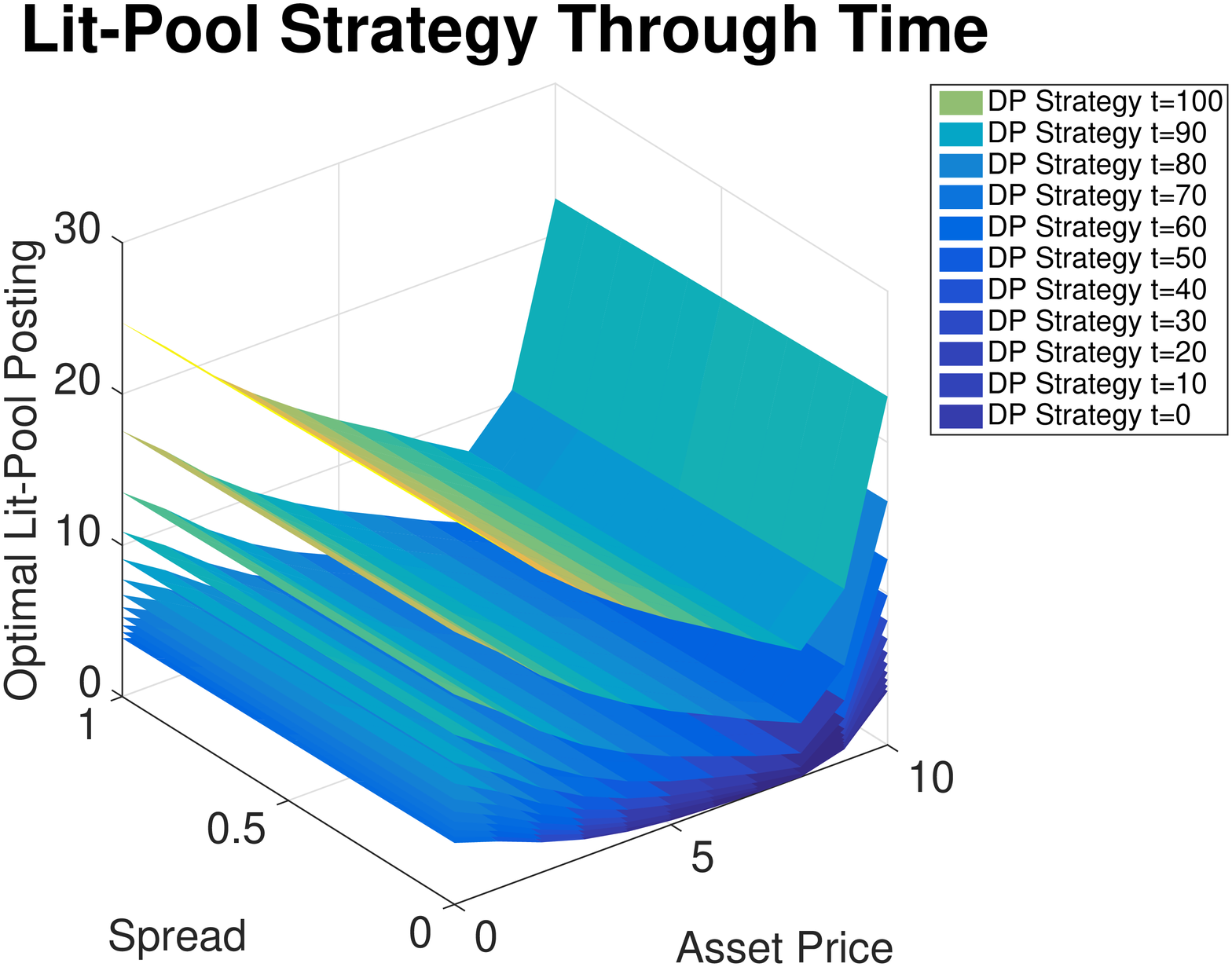}
\includegraphics[width=190pt,height=125pt]{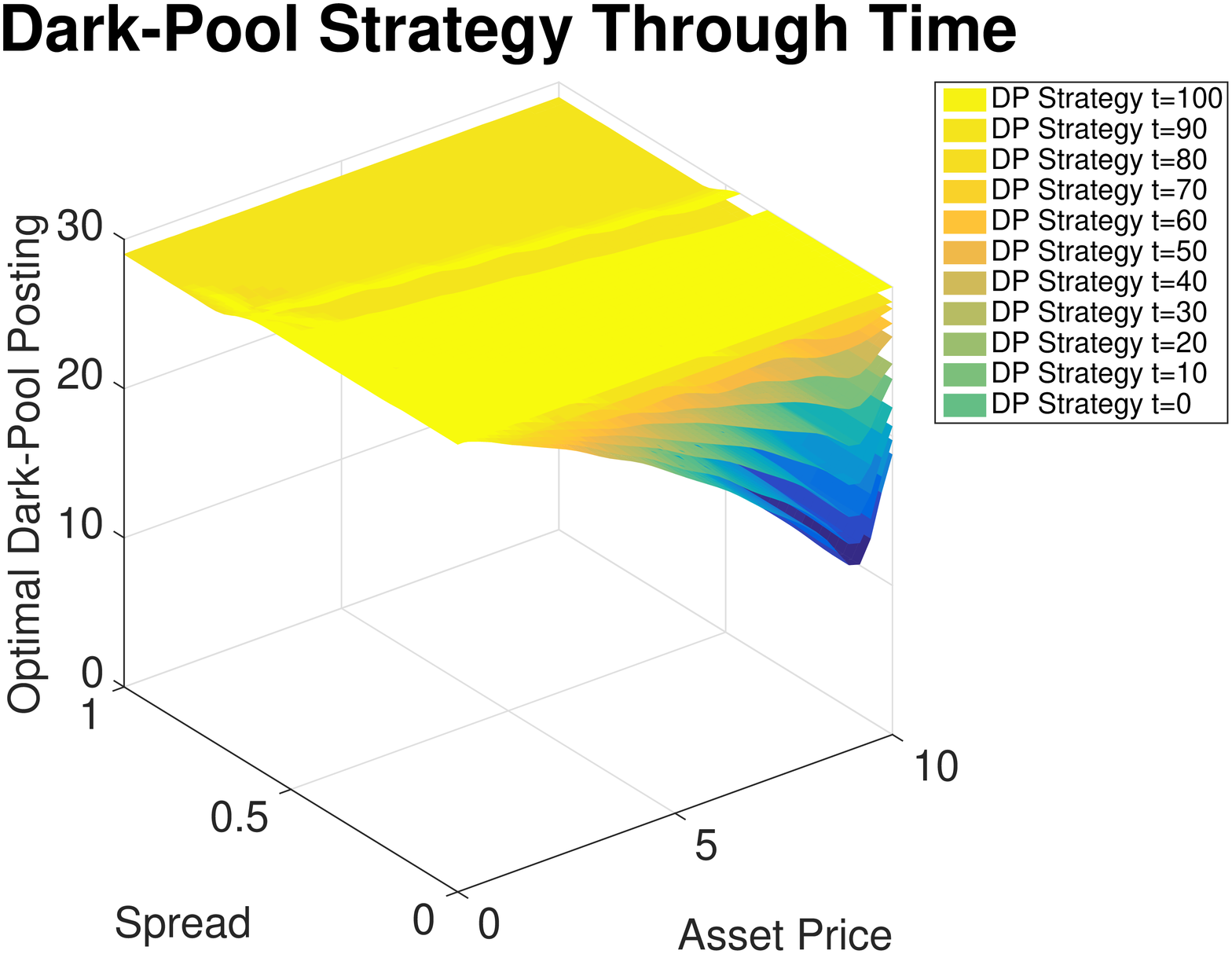}
\includegraphics[width=190pt,height=125pt]{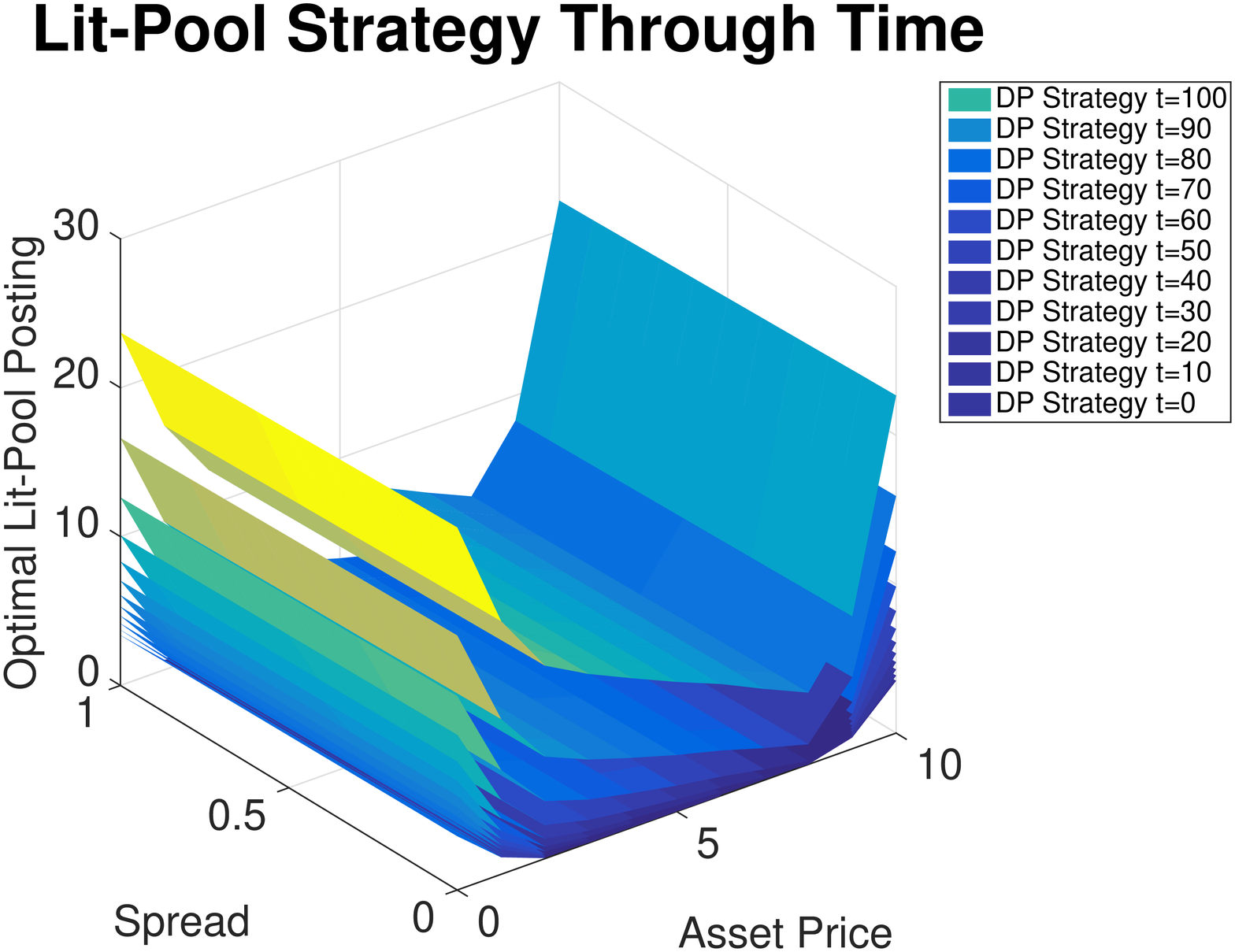}
\includegraphics[width=190pt,height=125pt]{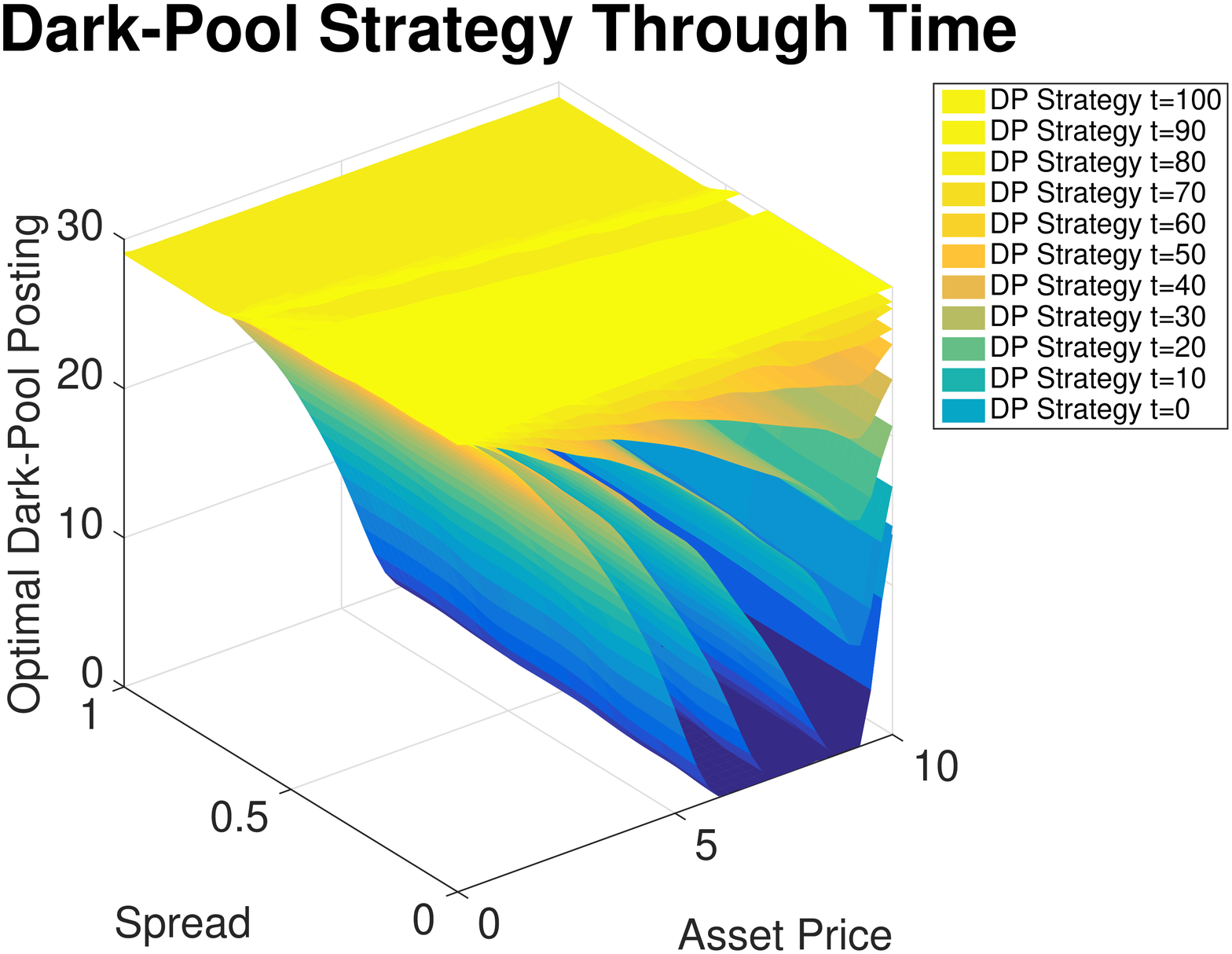}
\caption{\footnotesize{Optimal lit and dark pool strategies. We set  $X_t=30,000$ for each $t$, $\alpha=2$, $\gamma=0.0001$, $\lambda_1^b=0.1$, $\lambda_2^b=0.5$, $\lambda^{\Delta}_1=\lambda^{\Delta}_2=0.2$, $z_{i},w_{i}\sim U[0,0.1]$, $\beta=1\times10^{-5}$, $z^{w}\sim U[0,1]$, $\lambda^{w}=0.1$, $\mu^\Delta=\mu^b=0.001$. Top panels: $\lambda_1^b=\lambda_2^\Delta=0.5$, $\lambda^{b}_2=\lambda^{\Delta}_1=0.1$. Bottom panels: $\lambda_1^b=\lambda_2^\Delta=0.1$,  $\lambda^{b}_2=\lambda^{\Delta}_1=0.5$.
}}\label{eq:roundd}
\end{figure}

We propose next an exponential model for the best bid price and the spread process, so to ensure their positivity at every point in time, while upholding all other assumptions made in Section \ref{eq:meanr}. In particular, we set
\begin{align}
\label{eq:bid2}\frac{ \rd S^b\left(u\right)}{S^{b}(u^{-})}=&\ -\mu^{b}\nu\,(u)  \rd u+\rd J_1^{b}(u)-\rd J_2^{b}(u),\\
 \label{eq:ask2}\frac{ \rd \Delta\left(u\right)}{\Delta(u^{-})}=&\ \ +\mu^{\Delta}\nu\,(u) \rd u+\rd J_1^{\Delta}(u)-\rd J_2^{\Delta}(u)-\rd J_1^{b}(u)+\rd J_2^{b}(u).
\end{align}
\noindent Here, $\mu^{b}$ and $\mu^{\Delta}$ are the coefficients of the permanent price impact caused by the lit pool orders submitted by the agent. All other quantities are defined as in the previous example. In  Figure \ref{eq:gr5},  we plot a simulation of the best ask, the mid and the best bid prices.
By considering the same problem treated in Section \ref{eq:meanr}, Equation (\ref{eq:probl}), the associated  HJB PIDE is given by
\begin{equation*}
\begin{split}
\sup_{\boldsymbol{v}\in\mathcal Z}\,\biggl\{-\gamma x^2\!+\frac{\partial V}{\partial t}\!\left(t,\boldsymbol{x}\right)-\mu^{b}v s^{b}\frac{\partial V}{\partial s^{b}}\!\left(t,\boldsymbol{x}\right)+\mu^{\Delta}v \Delta\frac{\partial V}{\partial \Delta}\!\left(t,\boldsymbol{x}\right)+v(s^{b}-\beta v)\frac{\partial V}{\partial w}\!\left(t,\boldsymbol{x}\right)&\\
-v\frac{\partial V}{\partial x}\!\left(t,\boldsymbol{x}\right)+\lambda^{w}\E\left[V\left(t,x-n z^{w},s^{b},\Delta, w+n z^{w}\left(s^{b}+\frac{\Delta}{2}\right)   \right)\!-\!V\left(t,\boldsymbol{x}\right)\right]&\\
+\lambda_1^{b}\E\Big[V\!\left(t,x,s^{b}(1+z_{1}),\Delta(1-z_{1}), w\right)\!-\!V\!\left(t,\boldsymbol{x}\right)\Big]&\\
+\!\lambda_2^{b}\E\Big[V\!\left(t,x,s^{b}(1-z_{2}),\Delta(1+z_{2}), w\right)\!-\!V\!\left(t,\boldsymbol{x}\right)\Big]&\\
+\lambda_1^{\Delta}\E\Big[V\!\left(t,x,s^{b}\!,\Delta(1+y_{1}), w\right)\!-\!V\!\left(t,\boldsymbol{x}\right)\Big]&\\
+\!\lambda_2^{\Delta}\E\Big[V\!\left(t,x,s^{b}\!,\Delta(1-y_{2}), w\right)\!-\!V\!\left(t,\boldsymbol{x}\right)\Big] &\biggr\}=0
\end{split}
\end{equation*}
\noindent with terminal condition $V\big(T,x,s^b,\Delta,w\big)=w+({s}^b-\alpha x)x$.  In Figure \ref{eq:ag555} we show the inventory evolution for the case of a risk-neutral and of a risk-averse investor (top panels). Furthermore, we show the optimal trading trajectories for an increasing and a decreasing bid price (bottom panels). We notice that an increasing bid price (i.e. $\lambda_1^b>\lambda_2^b$) causes the agent to decelerate trading in the standard exchange so to benefit from the future favourable market trend. The opposite holds for a decreasing bid price. As in Section \ref{eq:meanr}, we show the lit and dark pools full trading strategies as a function of the best bid price and the spread. In Figure \ref{eq:roundd} we show both, the case of an increasing bid price (top panels) and the case of a decreasing bid price (bottom panels). 

\begin{figure}[H]
\centering
\includegraphics[width=205pt,height=170pt]{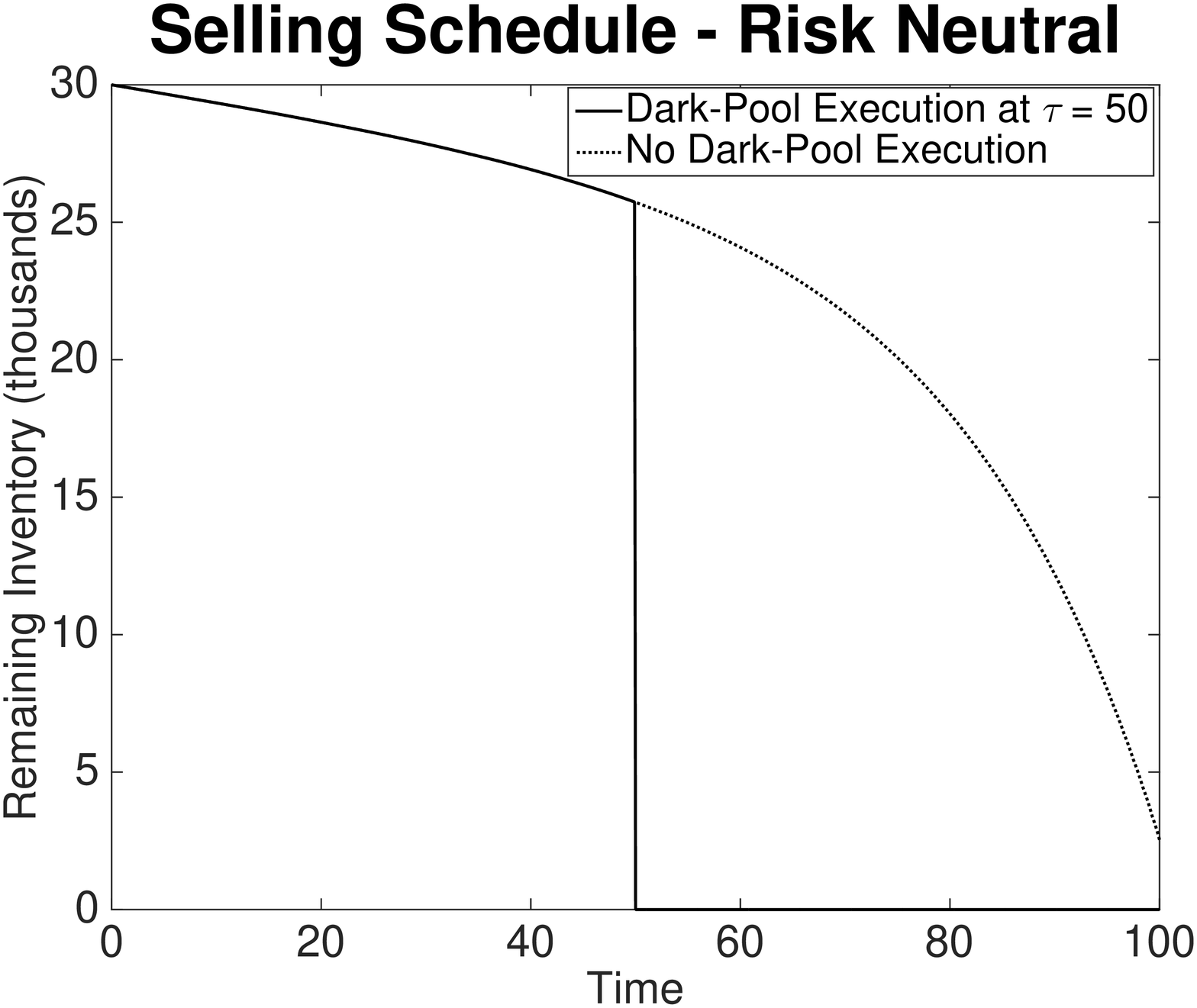}
\includegraphics[width=205pt,height=170pt]{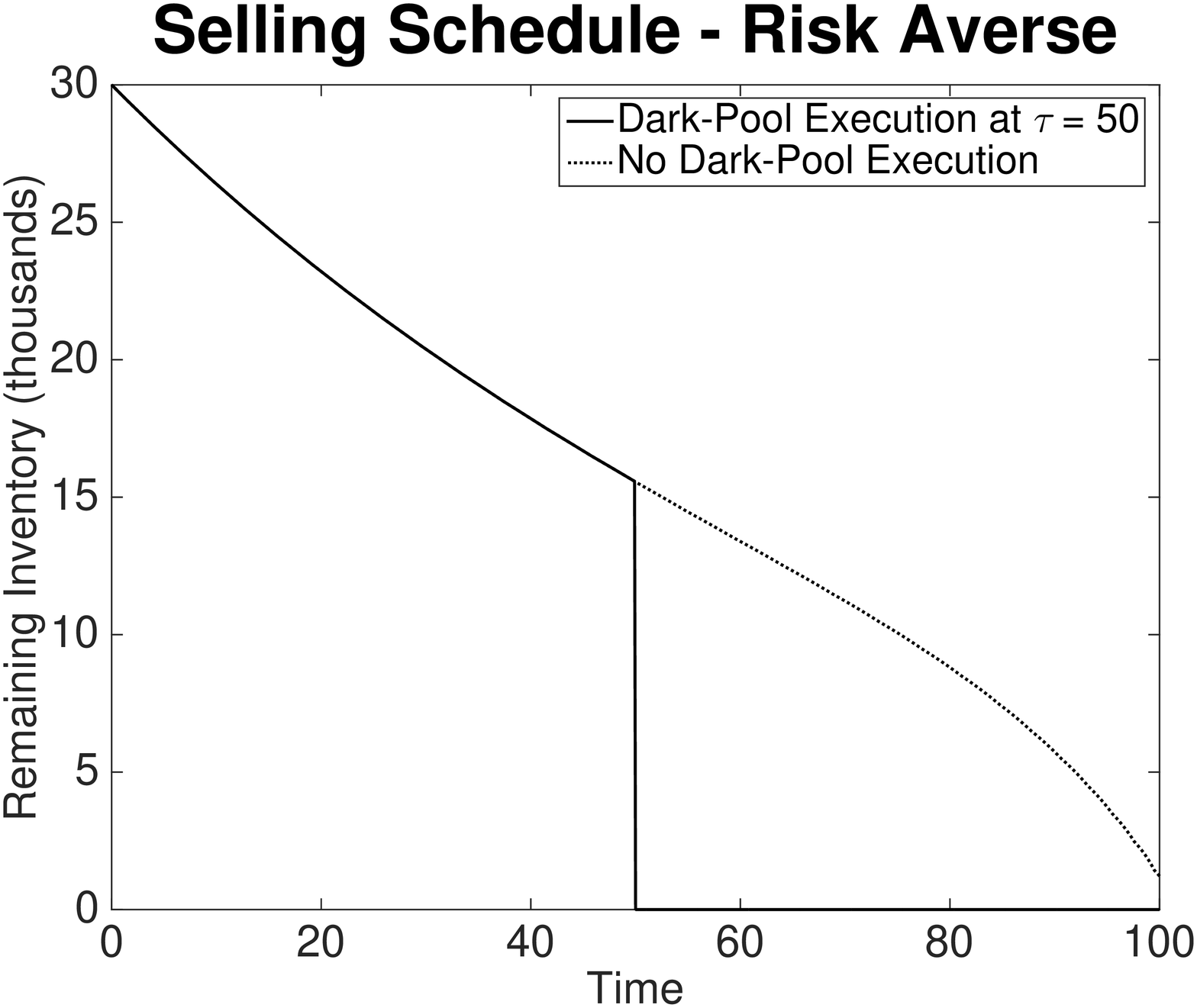}
\caption{\footnotesize{Optimal selling strategy displayed as a function of the remaining inventory. We set $\lambda_1^b=\lambda_2^b=\lambda^{\Delta}_1=\lambda^{\Delta}_2=0.2$, $z_{i},y_{i}\sim U[0,0.1]$, $s^{b}=40$, $\Delta\equiv0$, $\kappa^{b}=\mu^{b}=0$, $\beta=\alpha=1\times10^{-5}$. Left panel $\gamma=0$, and right panel  $\gamma=0.01$.
}}
\label{eq:krsh}
\end{figure}

For Figures \ref{eq:ag555} and \ref{eq:roundd} similar considerations as for the mean-reverting model apply.
The Kratz and Sch\"{o}neborn \cite{21} optimal trading strategy in the single-asset case, can be recovered in the present framework by
\begin{description}
\item[(a)] modelling $\{S^{b}(u)\}$ by a (local) martingale,
\item[(b)] setting  $\Delta(u)\equiv0\ \ \forall\ u\in[t,T]$,
\item[(c)] and setting the random variables $z^{w}_{i}\equiv1$ so to avoid partial execution.
\end{description}
In Figure \ref{eq:krsh}, we provide the results for the above choices in the case of a risk-neutral investor, $\gamma=0$, (left picture) and of a risk-averse investor, $\gamma>0$, (right picture). The shape of the policy shares the same features of the results obtained by Kratz and Sch\"{o}neborn \cite{21} (see Figure 4.1, page 17) without the constraint that total liquidation has to be achieved by $T$.

\section{Conclusions}\label{Sect.Conc}
In this paper we consider an optimal trading problem in the situation where an investor has access to both, a standard exchange and a dark pool. We propose novel continuous-time dynamics for the best bid price and the market spread, thus obtaining the mid-price, in an order-driven market, and capture the impact of the market activity in the Poisson random measures.  In particular we show how the best bid price changes due to the submission of limit buy orders at a more advantageous price, to the incoming market sell orders which may walk the book, or due to cancellations of limit sell orders posted at the best bid price. The orders in the dark pool are executed at the current mid-price in line with the available liquidity. We consider a partial execution setting so to account for various liquidity levels in the dark pool. We observe that the LOB dynamics plays an important role in determining the optimal trading strategy, for both the lit ad the dark pool. This is expected, for the price traded in the dark pool is taken from the lit pool. In fact, the mid-price  is influenced by the LOB dynamics, and so must be the optimal orders placement in the dark pool. We obtain the optimal trading strategies first in the case that only a standard exchange is available, and then also for when one may trade simultaneously in a dark pool. We discuss in detail the cases in which the market price dynamics is modelled by (i) a mean-reverting jump process and (ii) a geometric L\'evy process, and pay particular attention to properties usually less emphasised in the literature, e.g.  price dynamics which model realistic market features, the permanent price impact in simultaneous lit and dark pool trading strategies, and the effect of the sell side of the book in a liquidation problem. We employ the dynamic programming approach to obtain the optimal trading policies and we characterise---by means of standard arguments---the value function by the unique viscosity solution of the HJB PIDE associated to the optimal control problem. Although we treat a more general problem in Section \ref{eq:dp}, the considered models in Section \ref{eq:numerics} are closer to the classical optimal liquidation settings, making comparisons between the models more informative. The detailed model examples are also utilised to show that so-called round-trips, whereby an agent decides to post all the remaining inventory in the dark pool, are not generally beneficial. The discussion extends to the behaviour of market agents who have various levels of risk-appetite and who will find different strategies of simultaneous trading in lit and dark pools more attractive.
%%%%%%%%%%%%%%%%%%%%%%%%%%%%%%%%%%%%%%%%%%%%%%%%%%%%%%%%%%%%%%%%%%%%%%%%%%%%%%%%%%%%%%%%%%%%%%%%%%%%%%%%%%%%%%%%%%%%%%%%%%%%%%%%%%%%%%%%%%%%%%%%%%%%%%%%%%%%%%%%%%%%%%%%%%%%%%%%%%%%%%%%%%%%%%%%%%%%%%%%%%%%%%%%%%%%%%%%%%%%%%%%%%%%%%%%%%%%%%%%%%%%%%%%%%%%%%%%%%%%%%%%%%%%%%%%%%%%%%%%%%%%%%%%%%%%%%%%%%%%%%%%%%%%%%%%%%%%%%%%%%%%%%%%%%%%%%%%%%%%%%%%%%%%%%%%%%%%%%%%%%%%%%%%%%%%%%%%%%%%%%%%%%%%%%%%%%%%%%%%%%%%%%%%%%%%%%%%%%%%
\section{Appendix}
\subsection{Dark pool optimal control problem}\label{app-dp}
We give the details of the optimal control problem treated in Section \ref{eq:dp}. For $\boldsymbol{p}\in\RR^4$ with components $(p_1,p_2,p_3,p_{4})$, we define the operators ${\mathcal H}_1$, $\mathcal B_{b,\Delta}$ and $\mathcal B_{\Delta}$ by
\begin{equation*}
\begin{split}
{\mathcal H}_1\left(t,\boldsymbol{ x},\boldsymbol{ p}\right)=\sup_{{v}\in\mathcal V}\bigg\{\!f_1\left(t,\boldsymbol{x},{v}\right)-vp_1+\mu^b\left(t,s^b,v\right)p_2+\mu^{w}(t,s^{b},v)p_{3}+\mu\left(t,\Delta,v\right)p_4 \bigg\},
\end{split}
\end{equation*}
\begin{equation}
\begin{split}
\mathcal B_{b,\Delta}\left(t,\boldsymbol{x},\varphi\right)=\sum_{i=1}^{2}\int_\RR\bigg[\varphi\big(t,x,s^b+ h^{b}_i\big(t,s^b,z_{i}\big),\Delta +h^{b,\Delta}_i\big(t,s^b,\Delta,z_{i}\big),y    \big)-\varphi\big(t,\boldsymbol{x}\big)&\\
- h^{b}_i\big(t,s^b,z_{i}\big)\frac{\partial\varphi}{\partial{s^{b}}}\big(t,\boldsymbol{x}\big)- h^{b,\Delta}_i\big(t,s^b,\Delta,z_{i}\big)\frac{\partial\varphi}{\partial{\Delta}}\big(t,\boldsymbol{x}\big)\bigg]\Pi^{{b}}_{i}\big(\rd z_{i}\big)&,
\label{eq:909090}
\end{split}
\end{equation}
and 
\begin{equation}
\begin{split}
\mathcal B_{\Delta}\left(t,\boldsymbol{x},\varphi\right)=\sum_{i=1}^{2}\int_\RR\bigg[\varphi\big(t,x,s^b,\Delta + h_i\big(t,\Delta,y_{i}\big),w   \big)-\varphi\big(t,\boldsymbol{x}\big)&\\
- h_i\big(t,\Delta,y_{i}\big)\frac{\partial\varphi}{\partial{\Delta}}\big(t,\boldsymbol{x}\big)\bigg]&\Pi_{i}\big(\rd y_{i}\big).
\label{eq:808080}
\end{split}
\end{equation}
We also define the operator $\mathcal B_{w}$ by
\begin{equation}
\mathcal B_{w}(t,\boldsymbol{x},\varphi)=\sup_{{n}\in\mathcal N}\lambda^{w}\E\Big[\varphi(t,x-n z^{w},s^{b},\Delta,w+h^{w}(t,s^{b},\Delta,n)\,z^{w})-\varphi(t,\boldsymbol{x})\Big].
\end{equation}
Standard dynamic programming arguments suggest that the HJB equation associated to the optimisation problem in (\ref{eq:valuefunction1}) is a PIDE of the form
\begin{equation}
\begin{split}
rV\left(t,\boldsymbol{ x}\right)-\frac{\partial V}{\partial t}\left(t,\boldsymbol{ x}\right)-\mathcal H_1 \left(t,\boldsymbol{x},D_{\boldsymbol{ x}}V\right)-\mathcal B_{b,\Delta}(t,\boldsymbol{x},V)-\mathcal B_{\Delta}(t,\boldsymbol{x},V)-\mathcal B_{w}(t,\boldsymbol{x},V)=0,
\label{eq:pde1}
\end{split}
\end{equation}
on $[0,T)\times\mathcal O$, with terminal condition $V(T,\boldsymbol{x})=g_{1}(\boldsymbol{x})$. 
\begin{defn}
{\itshape 
A continuous function $V:[0,T)\times\mathcal O\rightarrow\RR$ is a viscosity subsolution (resp. supersolution) of the HJB Equation (\ref{eq:pde1}) if
\begin{equation*}
\begin{split}
r\phi\left(\bar t,\boldsymbol{\bar x}\right)-\frac{\partial \phi}{\partial t}\left(\bar t,\boldsymbol{\bar x}\right)\!-\!\mathcal H_1 \left(\bar t,\boldsymbol{\bar x},D_{\boldsymbol{ x}}\phi\right)\!-\!\mathcal B_{b,\Delta}(\bar t,\boldsymbol{\bar x},\phi)\!-\!\mathcal B_{\Delta}(\bar t,\boldsymbol{\bar x},\phi)\!-\!\mathcal B_{w}(\bar t,\boldsymbol{\bar x},\phi)&\leq 0
\end{split}
\end{equation*}
(resp. $\geq0$) for each $\phi\in \mathcal C^{1,1}([0,T)\times\mathcal O)\cap PB$ such that $V(t,\boldsymbol{x})-\phi(t,\boldsymbol{ x})$ attains its maximum (resp. minimum) at $(\bar{t},\boldsymbol{\bar x})\subset[0,T)\times\mathcal O$. A  continuous function is a viscosity solution if it is both a viscosity subsolution and a viscosity supersolution.
}
\end{defn}
\subsection{Assumptions}
We next provide technical assumptions underlying the models for the price (\ref{eq:bid}), the spread (\ref{eq:ask}) and cash account (\ref{eq:cash}) models, and we also state the assumptions for the running gains and terminal rewards in the value functions presented in Sections \ref{eq:exchange} and Appendix \ref{app-dp}.  

For $C,M>0$, $i=1,2$ and functions $\rho_{1,2}:{\RR}\rightarrow \RR^+$;   such that
\begin{equation}
\begin{split}
&\int_{\RR}\rho_1^2\big(y_i\big)\Pi_i\big(\rd y_i\big)<M\ \ ,\ \ \int_{\RR}\rho_2^2\big(z_i\big)\Pi^{b}_i\big(\rd z_i\big)<M,
\end{split}
\end{equation}
we assume that\footnote{We denote the $L_p$-norm by $\|\cdot\|_p$}
\begin{enumerate}[(i)]
 \item $\mu:[0,T]\times\RR_{+}\times \mathcal V\rightarrow \RR$ satisfies the Lipschitz continuity  and the linear growth conditions
\begin{equation}
\begin{split}
\big\vert {\mu}\left(t_{1},\Delta_{1},v\right)-{\mu}\left(t_{2},\Delta_2,v\right)\big\vert &\leq C\left(|t_{1}-t_{2}|+\big\vert \Delta_{1}-\Delta_2\big\vert \right),\\
 \big\vert {\mu}^b\left(t_{1},\Delta_{1},v\right)\big\vert &\leq C\left(1+\big\vert \Delta_{1}\big\vert \right),
 \end{split}
\end{equation}
for all $t_{1}$, $t_{2}\in[0,T]$, $\Delta_{1}$, $\Delta_2\in\RR_{+}$ and $v\in\mathcal V$;
 \label{eq:mmm1}

\item $\mu^b:[0,T]\times\RR_{+} \times\mathcal V\rightarrow \RR$ satisfies the Lipschitz continuity and the linear growth conditions
\begin{equation}
\begin{split}
\big\vert {\mu}^b\left(t_{1},s^b_{1},v\right)-{\mu}^b\left(t_{2},s^b_2,v\right)\big\vert &\leq C\left(|t_{1}-t_{2}|+\big\vert s^b_{1}-s^b_2\big\vert \right),\\
 \big\vert {\mu}^b\left(t_{1},s^b_{1},v\right)\big\vert &\leq C\left(1+\big\vert s^b_{1}\big\vert \right),
 \end{split}
\end{equation}
for all $t_{1}$, $t_{2}\in[0,T]$, $s^b_{1}$, $s^b_2\in\RR_{+}$ and $v\in\mathcal V$;
 \label{eq:mmmbbb}

\item $\mu^w:[0,T]\times\RR_{+} \times\mathcal V\rightarrow \RR$ satisfies the Lipschitz continuity and the linear growth conditions
\begin{equation}
\begin{split}
\big\vert {\mu}^w\left(t_{1},s^b_{1},v\right)-{\mu}^w\left(t_{2},s^b_2,v\right)\big\vert &\leq C\left(|t_{1}-t_{2}|+\big\vert s^b_{1}-s^b_2\big\vert \right),\\
 \big\vert {\mu}^w\left(t_{1},s^b_{1},v\right)\big\vert &\leq C\left(1+\big\vert s^b_{1}\big\vert \right),
 \end{split}
\end{equation}
for all $t_{1}$, $t_{2}\in[0,T]$, $s^b_{1}$, $s^b_2\in\RR_{+}$ and $v\in\mathcal V$;

\item for $h^{b,\Delta}_i:[0,T]\times\RR_{+}\times\RR_{+}\times\RR\rightarrow \RR$ is uniformly bounded for $|y_{i}|<1$ and satisfies the Lipschitz continuity and the linear growth conditions
\begin{equation}
\begin{split}
\big\vert h^{b,\Delta}_i\left(t_{1},s^b_1,\Delta_{1},y_{i}\right)-h^{b,\Delta}_i\left(t_{2},s^b_2,\Delta_{2},y_{i}\right)\big\vert &\leq \rho_1\left(y_i\right)\left(\big\vert t_{1}-t_2\big\vert+\big\vert s^b_{1}-s^b_2\big\vert+\big\vert \Delta_{1}-\Delta_2\big\vert \right),\\
 \big\vert h^{b,\Delta}_i\left(t_{1},s^b_1,\Delta_{1},y_{i}\right)\big\vert &\leq \rho_1\left(y_{i}\right)\left(1+\big\vert s^b_{1}\big\vert+\big\vert t_{1}\big\vert+\big\vert \Delta_{1}\big\vert \right),
 \end{split}
\end{equation}
 for all  $t_{1}$, $t_{2}\in[0,T]$, $s^b_{1}$, $s^b_2 \in\RR_{+}$ and $\Delta_{1}$, $\Delta_2\in\RR_{+}$.
  \label{eq:hhh}
  
  \item for $h_i:[0,T]\times\RR_{+}\times\RR\rightarrow \RR$ is uniformly bounded for $|y_{i}|<1$ and satisfies the Lipschitz continuity and the linear growth conditions
\begin{equation}
\begin{split}
\big\vert h_i\left(t_{1},\Delta_{1},y_{i}\right)-h_i\left(t_{2},\Delta_2,y_{i}\right)\big\vert &\leq \rho_1\left(y_i\right)\left(\big\vert t_{1}-t_2\big\vert+\big\vert \Delta_{1}-\Delta_2\big\vert \right),\\
 \big\vert h_i\left(t_{1},\Delta_{1},y_{i}\right)\big\vert &\leq \rho_1\left(y_{i}\right)\left(1+\big\vert t_{1}\big\vert+\big\vert \Delta_{1}\big\vert \right),
 \end{split}
\end{equation}
 for all  $t_{1}$, $t_{2}\in[0,T]$ and $\Delta_{1}$, $\Delta_2\in\RR_{+}$.
  \label{eq:hhhdbdb}

 \item for $h^{b}_i:[0,T]\times\RR_{+}\times\RR\rightarrow \RR$ is uniformly bounded for $|z_{i}|<1$ and satisfies the Lipschitz continuity and the linear growth conditions
\begin{equation}
\begin{split}
\big\vert h^{b}_i\left(t_{1},s^{b}_{1},z_{i}\right)-h^{b}_i\left(t_{2},s^{b}_2,z_{i}\right)\big\vert &\leq \rho_2\left(z_i\right)\left(\big\vert t_{1}-t_2\big\vert+\big\vert s^{b}_{1}-s^{b}_2\big\vert \right),\\
 \big\vert h^{b}_i\left(t_{1},s^{b}_{1},z_{i}\right)\big\vert &\leq \rho_2\left(z_{i}\right)\left(1+\big\vert t_{1}\big\vert+\big\vert s^{b}_{1}\big\vert \right),
 \end{split}
\end{equation}
 for all  $t_{1}$, $t_{2}\in[0,T]$ and $s^{b}_{1}$, $s^{b}_2\in\RR_{+}$.
  \label{eq:hhhbbb}

  \item for $h^{w}:[0,T]\times\RR^{2}_{+}\times\mathcal N\rightarrow \RR$  satisfies the Lipschitz continuity and the linear growth conditions
\begin{equation}
\begin{split}
\big\vert h^{w}\left(t_{1},s^{b}_{1},\Delta_{1},n\right)-h^{w}\left(t_{2},s^{b}_{2},\Delta_{2},n\right)\big\vert &\leq C\left(\big\vert t_{1}-t_2\big\vert+\big\vert s^{b}_{1}-s^{b}_2\big\vert +\big\vert \Delta_{1}-\Delta_2\big\vert\right),\\
 \big\vert h^{w}\left(t_{1},s^{b}_{1},\Delta_{1},n\right)\big\vert &\leq C\left(1+\big\vert t_{1}\big\vert+\big\vert s^{b}_{1}\big\vert +\big\vert \Delta_{1}\big\vert \right),
 \end{split}
\end{equation}
 for all  $t_{1}$, $t_{2}\in[0,T]$, $n\in\mathcal N$ and $s^{b}_{1}$, $s^{b}_2$, $\Delta_{1}$, $\Delta_{2}\in\RR_{+}$.
 
  \label{eq:hhh3}

\item  the map $f:[0,T]\times\mathcal O\times\mathcal V\rightarrow \RR$ satisfies the Lipschitz continuity and the linear growth conditions\begin{equation}
\begin{split}
\big\vert f\left(t_{1},\boldsymbol{x}_{1},v\right)-f\left(t_{2},\boldsymbol{ x}_{2},v\right)\big\vert &\leq C\left(|t_{1}-t_{2}|+\|\boldsymbol{x}_{1}-\boldsymbol{x}_2\|_1\right),\\
 \big\vert f\left(t_{1},\boldsymbol{x}_{1},v\right)\big\vert &\leq C\left(1+\|\boldsymbol{x}_{1}\|_1\right),
 \end{split}
\end{equation}
for all $t_{1}$, $t_{2}\in[0,T]$, $\boldsymbol{x}_{1}$, $\boldsymbol{ x}_{2}\in\mathcal O$ and $v\in\mathcal V$;
 \item $g:\mathcal O\rightarrow \RR$ satisfies the Lipschitz continuity  
 \begin{equation}
\big\vert g\left(\boldsymbol{x}_{1}\right)-g\left(\boldsymbol{x}_{2}\right)\big\vert \leq C\|\boldsymbol{x}_{1}-\boldsymbol{x}_{2}\|_{1},
\end{equation}
for all  $\boldsymbol{x}_{1}$, $\boldsymbol{ x}_{2}\in\mathcal O$;
 
  \item  the map $f_1:[0,T]\times\mathcal O\times\mathcal V\rightarrow \RR$ satisfies the Lipschitz continuity and the linear growth conditions
  \begin{equation}
\begin{split}
\big\vert f_1\left(t_{1},\boldsymbol{x}_{1},{v}\right)-f_1\left(t_{2},\boldsymbol{ x}_2,{ v}\right)\big\vert &\leq C\left(\vert t_{1}-t_{2}\vert +\|\boldsymbol{x}_{1}-\boldsymbol{x}_2\|_1\right),\\
 \big\vert f_1\left(t_{1},\boldsymbol{x}_{1},{ v}\right)\big\vert &\leq C\left(1+\|\boldsymbol{x}_{1}\|_1\right),
 \label{eq:f1}
 \end{split}
\end{equation}
for all $t_{1}$, $t_{2}\in[0,T]$, $\boldsymbol{x}_{1}$,  $\boldsymbol{ x}_{2}\in\mathcal O$ and ${ v} \in\mathcal V$;
 \item  $g_1:\mathcal O\rightarrow \RR$  satisfies the Lipschitz continuity  
 \begin{equation}
\big\vert g_{1}\left(\boldsymbol{x}_{1}\right)-g_{1}\left(\boldsymbol{x}_{2}\right)\big\vert \leq C\|\boldsymbol{x}_{1}-\boldsymbol{x}_{2}\|_{1},
\end{equation}
for all  $\boldsymbol{x}_{1}$, $\boldsymbol{ x}_{2}\in\mathcal O$; 
\end{enumerate}
Standard results---see, e.g., Ikeda and Watanabe \cite{19}---ensure that under the assumptions (\ref{eq:mmm1}) to (\ref{eq:hhh3}) there exists a strong and path-wise unique solution of the price, the spread and the wealth models defined by Equations (\ref{eq:bid}), (\ref{eq:ask}) and (\ref{eq:cash}). 
\subsection{Moments Estimates}
We now provide some moments estimates of  $S^b(u)$. We note that both, the following proposition and its proof, are analogous for $\Delta(u)$, $X(u)$ and $W(u)$.

\begin{prop}
{\itshape Fix   $p=1,2$. Let the assumptions (\ref{eq:mmmbbb}) and (\ref{eq:hhhbbb})  hold and let $S^b_{t,\,s^b}(u)$ be the random variable  at a fixed time $u\in[t,T]$ with initial values $(t,s^b)\in[0,T]\times \RR_{+}$. Then, for any  $v\in\mathcal V$ and for any stopping time $\tau_0\leq h\in[0,T]$ there exists a constant $K=K(p,C,M,T)>0$ such that
\begin{equation}
\begin{split}
\E\left[\,\big\vert \,S^b_{t,\,s^b}(\tau_0)\,\big|^p\right]&\leq K\left(1+\,\big\vert \,s^b\,\big\vert^p\right),\\
\E\left[\,\big\vert \,S^b_{t,\,s^b_{1}}(\tau_0)-S^b_{t,\,s^b_2}(\tau_0)\,\big\vert^p\right]&\leq K\left(\,\big\vert \,s^b_{1}-s^b_2\,\big\vert^p\right),\\
\E\left[\,\big\vert \,S^b_{t,\,s^b}(\tau_0)-s^b\,\big\vert^p\right]&\leq K\left(1+\,\big\vert \,s^b\,\big\vert^p\right)(h-t)^\frac{p}{2},\\
\E\left[\,\sup_{0\leq u\leq h}\big\vert \,S^b_{t,\,s^b}(u)-s^b\,\big\vert^p\right]&\leq K\left(1+\,\big\vert \,s^b\,\big\vert^p\right)(h-t)^\frac{p}{2}.
\label{eq:3666}
\end{split}
\end{equation}
\label{eq:momentsa}}
\end{prop}
\noindent{\itshape Proof.} We adapt the proof in  Pham \cite{27} to the present work and indeed we shall consider the proof only for $p=2$ as it  suffices to ensure the relation for $p=1$, according to H\"{o}lder's inequality. In order to reduce notation, here $K$ is a generic positive constant which may take different values in different places. Define  $\mathcal T_h$ as the set of all stopping times smaller than $h\in[0,T]$. By the optional sampling theorem and the L\'evy-It\^o isometry, we have 
\begin{equation*}
\begin{split}
&\E\left[\,\big\vert\,S^b_{t,\,s^b}(\tau_0)\,\big\vert^2\right]\leq\\
&K\E\left[\,\big\vert\,s^b\,\big\vert^2\!\!+\!\!\int_{t}^{\tau_0}\!\!\left|\,\mu^b\!\left(u,S^b_{t,\,s^b}(u),v(u)\right)\right|^2\rd u\!+\sum_{i=1}^2\!\int_{t_1}^{\tau_0}\!\!\int_{\RR}\!\left|\,h^{b}_i\!\left(u,S^b_{t, \,s^b}(u),z_i\right)\right|^2\Pi^{b}_i\left(\rd z_i\right)\rd u\right]\!\!,\\
\end{split}
\end{equation*}
for $\tau_0\in\mathcal T_h$. By the linear growth conditions on $\mu^b$, $ h^{b}_1$, $h^{b}_2$, we have
\begin{equation}
\E\left[\,\big\vert\,S^b_{t,\,s^b}(\tau_0)\,\big\vert^2\right]\leq K\left[  1+\,\big\vert\,s^b\big\vert\,^2 +\E\!\int_{t}^{\tau_0}\,\big\vert\,S^b_{t,\,s^b}(u)\,\big\vert^2\rd u\right].
\label{eq:1est}
\end{equation}
As noted in Pham (1998), if $\tau_0$ were a deterministic time, (\ref{eq:1est}) would yield
\begin{equation*}
\E\left[\,\big\vert\,S^b_{t,\,s^b}(\tau_0)\,\big\vert^2\right]\leq K\left[  1+\,\big\vert\,s^b\,\big\vert^2 \right].
\end{equation*}
By definition of $\mathcal T_h$, we note that
$$
\E\left[\int_{t}^{\tau_0}\,\big\vert\,S^b_{t,\,s^b}(u)\,\big\vert^2\rd u\right]\leq\E\left[\int_{t}^{h}\,\big\vert\,S^b_{t,\,s^b}(u)\,\big\vert^2\rd u\right].
$$
Thus, by applying Fubini's theorem to exchange the order of integration and by Gronwall's lemma,
\begin{equation}
\E\left[\,\big\vert\,S^b_{t,\,s^b}(\tau_0)\,\big\vert^2\right]\leq K\left[  1+\,\big\vert\,s^b\,\big\vert^2 +\!\int_{t}^{h}\E\left[\,\big\vert\,S^b_{t,\,s^b}(u)\,\big\vert^2\right]\rd u\right]\leq K\left[  1+\,\big\vert\,s^b\,\big\vert^2 \right],
\end{equation}
for a suitable constant $K=K(p,C,M,T)$. Define the process $Z(u)$ by
$$
Z(u)=S^b_{t,\,s^b_{1}}(u)-S^b_{t,\,s^b_2}(u).
$$
Then by an application of  It$\hat{\textrm{o}}$'s formula to $|Z(u)|^{2}$, we have\begin{equation*}
\begin{split}
\E\left[\left|Z(\tau_0)\right|^2\right]= \E&\left[\left|s^b_{1}-s^b_2\right|^2\!+\!\int_{t}^{\tau_0}\!\!\!\!2Z(u)\,\Big(\,\mu^b\Big(u,S^b_{t,\,s^b_{1}}(u),v(u)\Big)-\mu^b\Big(u,S^b_{t,\,s^b_2}(u),v(u)\Big)\Big)\rd u\right.\\
&\left.+\sum_{i=1}^2\!\int_{t}^{\tau_0}\!\int_{\RR} \Big|\,h^{b}_i\Big(u,S^b_{t,\,s^b_{1}}(u),z_i\Big)\!-\!h^{b}_i\Big(u,S^b_{t,\,s^b_2}(u),z_i\Big)\Big|^2\Pi_i^{b}\left(\rd z_i\right)\rd u\right].
\end{split}
\end{equation*}
From the Lipschitz condition on $\mu^b$, $ h^{b}_1$ and $ h^{b}_2$, it follows that
\begin{equation*}
\E\!\left[\,\big\vert\,Z(\tau_0)\,\big\vert^2\right]\!\leq\! K\E\!\left[\,\big\vert\,s^b_{1}-s^b_2\,\big\vert^2\!\!+\!\!\int_{t}^{\tau_0}\!\!\!\big\vert\,S^b_{t,\,s^b_{1}}(u)-S^b_{t,\,s^b_2}(u)\,\big\vert^2\rd u\right]\!.
\end{equation*}
By making use of Fubini's Theorem and Gronwall's Lemma we get
\begin{equation*}
\E\left[\,\big\vert\,S^b_{t,\,s^b_{1}}({\tau_0})-S^b_{t,\,s^b_2}({\tau_0})\,\big\vert^2\right]\leq\! K\E\!\left[\,\big\vert\,s^b_{1}-s^b_2\,\big\vert^2\!+\!\int_{t}^{h}\!\!\!\big|Z(u)\big|^{2}\rd u\right]\!\!\leq K\,\big\vert\,s^b_{1}-s^b_2\,\big\vert^2,
\end{equation*}
for a suitable constant $K=K(p,C,M,T)$. For the third moment estimate, we make use of the first moment estimate in (\ref{eq:3666}) to obtain
\begin{equation*}
\E\left[\,\big\vert\,S^b_{t,\,s^b}(t)-s^b\,\big\vert^2\right]\leq K\left[\int_{t}^{\tau_0}\!\!\!\left(1+\E\left[\,\big\vert\,S^b_{t,\,s^b}(u)\,\big\vert^2\right]\right)\rd u \right]\leq K\left(1+\,\big\vert\,s^b\,\big\vert^2\right)(h-t).
\end{equation*}
The fourth moment estimate in (\ref{eq:3666}) follows from the third moment estimate,  Doob's maximal inequality, and the fact that the constant $K$ does not depend on the control process.$\hfill\square$
\subsection{Viscosity Solution}
In what follows, we note that it suffices to show the viscosity property for the model presented in Section \ref{eq:dp}, since the model discussed in Section \ref{eq:optim} is a special case.

\begin{prop}
{\itshape The value function $V:[0,T]\times\mathcal O \rightarrow \RR$ defined in (\ref{eq:valuefunction1}) is continuous on $[0,T]\times\mathcal O$. Furthermore, for $K>0$ and $\forall\ \boldsymbol{x}\in\mathcal O$, it satisfies
\begin{equation}
V\left(t,\boldsymbol{x}\right)\leq K\left(1+\|\boldsymbol{x}\|_1\right).
\label{eq:lineargr}
\end{equation}
\label{eq:dpcont}}
\end{prop}
\noindent{\itshape Proof.}
We proceed in two steps. We first show that the value function is Lipschitz continuous in $\boldsymbol{x}$, uniformly in $t$. Next we show that it is continuous in $t$. We take $\boldsymbol{x},\boldsymbol{y}\in\mathcal O$ and  since $|\sup(A)-\sup(B)|\leq\sup|(A-B)|$, we have that
\begin{equation*}
\begin{split}
|V\!\left(t,\boldsymbol{x}\right)-V\!\left(t,\boldsymbol{y}\right)|=&\Bigg|\sup_{\boldsymbol{\nu}\in\mathcal Z}\E\!\left[\int_{t}^\tau\!\! \e^{-r(u-t)}f_1\!\left(u,\boldsymbol{ X}_{t,\,\boldsymbol{x}}(u),{\nu}(u)\right)\rd u+\e^{-r(\tau-t)}g_1\!\left(\boldsymbol{X}_{t,\,\boldsymbol{x}}(\tau)\right)\right]\\
&- \sup_{\boldsymbol{\nu}\in\mathcal Z}\E\!\left[\int_{t}^\tau\!\! \e^{-r(u-t)}f_1\!\left(u,\boldsymbol{ X}_{t,\,\boldsymbol{y}}(u),{\nu}(u)\right)\rd u+\e^{-r(\tau-t)}g_1\!\left(\boldsymbol{X}_{t,\,\boldsymbol{y}}(\tau)\right)\right]\Bigg|\\
\leq&\sup_{\boldsymbol{\nu}\in\mathcal Z}\Bigg|\E\Bigg[\int_{t}^\tau\!\! \e^{-r(u-t)}\Big(f_1\!\left(u,\boldsymbol{ X}_{t,\,\boldsymbol{x}}(u),{\nu}(u)\right)\rd u-f_1\!\left(u,\boldsymbol{ X}_{t,\,\boldsymbol{y}}(u),{\nu}(u)\right)\Big)\rd u\\
&+\e^{-r(\tau-t)}\Big(g_1\!\left(\boldsymbol{X}_{t,\,\boldsymbol{x}}(\tau)\right)-g_1\!\left(\boldsymbol{X}_{t,\,\boldsymbol{y}}(\tau)\right)\Big)\Bigg]\Bigg|.
\end{split}
\end{equation*}
The Lipschitz continuity of $f_{1}$ and $g_{1}$ give
\begin{equation*}
\begin{split}
|V\!\left(t,\boldsymbol{x}\right)-&V\!\left(t,\boldsymbol{y}\right)|\\
&\leq\sup_{\boldsymbol{\nu}\in\mathcal Z}\E\Bigg[\int_{t}^\tau\!\! \e^{-r(u-t)}K|\boldsymbol{ X}_{t,\,\boldsymbol{x}}(u)-\boldsymbol{ X}_{t,\,\boldsymbol{y}}(u)|\rd u+\e^{-r(\tau-t)}K|\boldsymbol{X}_{t,\,\boldsymbol{x}}(\tau)-\boldsymbol{X}_{t,\,\boldsymbol{y}}(\tau)|\Bigg]\\
&\leq K||\boldsymbol{x}-\boldsymbol{y}||_{1},
\end{split}
\end{equation*}
where the last inequality is justified by the moments estimates in Proposition \ref{eq:momentsa}. We note that an analogous calculation will produce Equation (\ref{eq:lineargr}). We now take $0\leq t_{1}<t_{2}<T$ and we apply the DP to obtain
\begin{equation*}
\begin{split}
|V\!\left(t_{1},\boldsymbol{x}\right)&-V\!\left(t_{2},\boldsymbol{x}\right)|=\Bigg|\sup_{\boldsymbol{\nu}\in\mathcal Z}\E\Bigg[\int_{t_{1}}^{t_{2}\wedge\tau}\!\! \!\!\!\!\!\e^{-r(u-t_{1})}f_1\!\left(u,\boldsymbol{ X}_{t_{1},\,\boldsymbol{x}}(u),{\nu}(u)\right)\rd u\\
&+\e^{-r(t_{2}-t_{1})}V\!\left(t_{2},\boldsymbol{X}_{t_{1},\,\boldsymbol{x}}(t_{2})\right)\mathbbm{1}_{\{\tau\geq t_{2}\}}+\e^{-r(\tau-t_{1})}g_{1}\!\left(\boldsymbol{X}_{t_{1},\,\boldsymbol{x}}(\tau)\right)\mathbbm{1}_{\{\tau<t_{2}\}}\Bigg]-V\!\left(t_{2},\boldsymbol{x}\right)\!\Bigg|.
\end{split}
\end{equation*}
We can add and subtract the quantity 
$$
\mathbbm{1}_{\{\tau<t_{2}\}}\e^{-r(\tau-t_{1})}(g_{1}(\boldsymbol{x})+V(t_{2},\boldsymbol{x}))+\mathbbm{1}_{\{\tau\geq t_{2}\}}\e^{-r(t_{2}-t_{1})}V(t_{2},\boldsymbol{x}),
$$
to obtain
\begin{equation*}
\begin{split}
&|V\!\left(t_{1},\boldsymbol{x}\right)-V\!\left(t_{2},\boldsymbol{x}\right)|\leq \\
&\sup_{\boldsymbol{\nu}\in\mathcal Z}\E\Bigg[\int_{t_{1}}^{t_{2}\wedge\tau}\!\! \!\!\!\!\!\e^{-r(u-t_{1})}|f_1\!\left(u,\boldsymbol{ X}_{t_{1},\,\boldsymbol{x}}(u),{\nu}(u)\right)|\rd u+ \mathbbm{1}_{\{\tau\geq t_{2}\}}\e^{-r(t_{2}-t_{1})}|V\!\left(t_{2},\boldsymbol{X}_{t_{1},\,\boldsymbol{x}}(t_{2})\right)-V(t_{2},\boldsymbol{x})|     \\
&+\mathbbm{1}_{\{\tau<t_{2}\}}\e^{-r(\tau-t_{1})}|g_{1}\!\left(\boldsymbol{X}_{t_{1},\,\boldsymbol{x}}(\tau)\right)-g_{1}(\boldsymbol{x})|+\mathbbm{1}_{\{\tau<t_{2}\}}\e^{-r(\tau-t_{1})}|  g_{1}(\boldsymbol{x})-V(t_{2},\boldsymbol{x}) | \\
& +\mathbbm{1}_{\{\tau<t_{2}\}}|(\e^{-r(\tau-t_{1})}-1)V\!\left(t_{2},\boldsymbol{x}\right)|+ \mathbbm{1}_{\{\tau\geq t_{2}\}}|(\e^{-r(t_{2}-t_{1})}-1)V\!\left(t_{2},\boldsymbol{x}\right)|\Bigg]\leq K|t_{2}-t_{1}|(1+|\boldsymbol{x}|),\\
\end{split}
\end{equation*}
where the last inequality is justified by: (i) the linear growth of $f_{1}$,  $g_{1}$ and $V$, the Lipschitz continuity of $g_{1}$ and of $V$ in $\boldsymbol{x}$ uniformly in $t$, and the moment estimates in Proposition \ref{eq:momentsa}.
Thus, we can conclude that
$$
|V\!\left(t_{1},\boldsymbol{x}\right)-V\!\left(t_{2},\boldsymbol{y}\right)|\leq K(|t_{2}-t_{1}|(1+|\boldsymbol{x}|)+||\boldsymbol{x}-\boldsymbol{y}||).
$$

\hfill$\square$

\begin{prop}
{\itshape The value function $V$ defined by Equation (\ref{eq:valuefunction1}) is a viscosity solution of the HJB PIDE (\ref{eq:pde1}).
\label{eq:dpvisc}}
\end{prop}
\noindent{\itshape Proof.} We show that $V(t,\boldsymbol{ x})$ is a continuous  viscosity solution of (\ref{eq:pde1}) by proving that it is both a supersolution and a subsolution. We proceed along the same lines of {\O}ksendal and Sulem \cite{oa} and we first show the supersolution property. We define a test function $\phi:[0,T)\times\mathcal O\rightarrow \RR$ such that $\phi\in \mathcal C^{1,1}([0,T)\times\mathcal O)\cap PB$ and, without loss of generality, we assume that $V-\phi$ reaches its minimum at $(\bar t,\boldsymbol{\bar x})$, such that
\begin{equation}
V\left(\bar{t},\boldsymbol{\bar x}\right)-\phi\left(\bar{t},\boldsymbol{\bar x}\right)=0.
\label{eq:min}
\end{equation}
We let $\tau_{1}$ be a stopping time defined by $\tau_{1}=\inf \,\{u>\bar t\,\vert\,\boldsymbol{X}_{\bar t,\, \boldsymbol{ \bar x}}(u)\notin B_{\epsilon}( \boldsymbol{\bar x})\}$, where $B_{\epsilon}( \boldsymbol{ \bar x})$ is the ball of radius $\epsilon$ centred in  $ \boldsymbol{ \bar x}$.  Then we define the stopping time $\tau^{*}=\tau_{1}\wedge(\bar{t}+h)$ for $0< h<T-\bar{t}$ and note that $\bar \gamma:=\E_{\bar t\,\boldsymbol{\bar x}}[\tau^{*}]>0$. 
From the first part of DP and the definition of  $\phi$, it follows that, for arbitrary $\boldsymbol{\nu}\in\mathcal Z$,
\begin{equation*}
\begin{split}
V\left(\bar{t},\boldsymbol{\bar{x}}\right)\geq \E&\!\left[\!\int_{\bar{t}}^{\tau^{*}}\!\!\! \e^{-r(u-\bar t)}f_1\left(u,\boldsymbol{X^{\nu}}_{\bar t,\,\boldsymbol{\bar x}}(u),{\nu}(u)\right)\!\rd u\!+\e^{-r(\tau^{*}-\bar{t})}\phi\left(\tau^{*}\!\!,\boldsymbol{X^{\nu}}_{\bar t,\,\boldsymbol{\bar x}}(\tau^*)\right) \right]\!\!.
\end{split}
\end{equation*}
By applying Dynkin's formula to $\e^{-r(\tau^{*}-\bar t)}\phi\big(\tau^{*}\!\!,\boldsymbol{ X^{\nu}}_{\bar t,\,\boldsymbol{\bar x}}(\tau^*)\big)$ at  $(\bar{t},\boldsymbol{\bar x})$,  we get
\begin{equation*}
\begin{split}
\E\Bigg[\!\int_{\bar{t}}^{\tau^{*}}\!\!\Biggl(&\e^{-r(u-\bar t)}f_1\left(u,\boldsymbol{ X^{\nu}}_{\bar t,\,\boldsymbol{\bar x}}(u),{\nu}(u)\right)-\e^{-r(u-\bar t)}\,\Biggl\{r\phi\left(u,\boldsymbol{ X^{\nu}}_{\bar t,\,\boldsymbol{\bar x}}(u)\right)\!- \frac{\partial \phi}{\partial t}\left(u,\boldsymbol{ X^{\nu}}_{\bar t,\,\boldsymbol{\bar x}}(u)\right)\\
&-\mathcal{\bar{H}}_1 \left(u,\boldsymbol{ X^{\nu}}_{\bar t,\,\boldsymbol{\bar x}}(u),D_{\boldsymbol{ x}}\phi,{\nu}(u)\right) - \mathcal B_{b,\Delta}(u,\boldsymbol{ X^{\nu}}_{\bar t,\,\boldsymbol{\bar x}}(u),\phi)-\mathcal B_{\Delta}(u,\boldsymbol{ X^{\nu}}_{\bar t,\,\boldsymbol{\bar x}}(u),\phi)\\
&-\bar{\mathcal B}_{w}(u,\boldsymbol{ X^{\nu}}_{\bar t,\,\boldsymbol{\bar x}}(u),\phi,\eta(u))\Biggr\}\Biggr)\rd u\Bigg]\leq0,
\end{split}
\end{equation*}
where $\mathcal{\bar{H}}_1$ and $\bar{\mathcal B}_{w}$ are defined respectively by
\begin{equation*}
\begin{split}
\bar{\mathcal  H}_1\left(t,\boldsymbol{ x},\boldsymbol{ p},{v}\right)=-vp_1+\mu^b\left(t,s^b,v\right)p_2+\mu^{w}(t,s^{b},v)p_{3}+\mu\left(t,\Delta,v\right)p_4 ,
\end{split}
\end{equation*}
and
\begin{equation*}
\begin{split}
\bar{\mathcal  B}_w(t,\boldsymbol{ x},\varphi,n)=\lambda^{w}\E\Big[\varphi(t,x-n z^{w},s^{b},\Delta,w+h^{w}(t,s^{b},\Delta,n)z^{w})-\varphi(t,\boldsymbol{x})\Big]. 
\end{split}
\end{equation*}
We  divide both sides by $-\bar \gamma$ and let $h\rightarrow 0$, resulting in
\begin{equation*}
\begin{split}
r\phi\left(\bar t,\boldsymbol{\bar x}\right)-\frac{\partial \phi}{\partial t}\left(\bar t,\boldsymbol{\bar x}\right)-\mathcal{\bar{H}}_1 \left(\bar t,\boldsymbol{\bar x},D_{\boldsymbol{x}}\phi,{v}\right)-f_1(\bar t,\boldsymbol{\bar x},{v})-\mathcal B_{b,\Delta}(\bar t,\boldsymbol{\bar x},\phi)-\mathcal B_{\Delta}(\bar t,\boldsymbol{\bar x},\phi)&\\
-\mathcal B_{w}(\bar t,\boldsymbol{\bar x},\phi,\eta)&\geq 0.
\end{split}
\end{equation*}
Due to the arbitrariness of $\boldsymbol{v}$, we can rewrite the above as
\begin{equation*}
r\phi\left(\bar t,\boldsymbol{\bar x}\right)-\frac{\partial \phi}{\partial t}\left(\bar t,\boldsymbol{\bar x}\right)-\mathcal H_1 \left(\bar t,\boldsymbol{\bar x},D_{\boldsymbol{ x}}\phi\right)-\mathcal B_{b,\Delta}(\bar t,\boldsymbol{\bar x},\phi)-\mathcal B_{\Delta}(\bar t,\boldsymbol{\bar x},\phi)-\mathcal B_{w}(\bar t,\boldsymbol{\bar x},\phi)\geq 0,
\end{equation*}
which proves the supersolution inequality. We now prove the subsolution inequality. We let $\phi$ be a smooth and polinomially-bounded test function  such that $V-\phi$ has its maximum at $(\bar{t},\boldsymbol{\bar x})$. Without loss of generality, we assume $V(\bar{t},\boldsymbol{\bar x})-\phi(\bar{t},\boldsymbol{\bar x})=0$. We shall show that the following inequality 
\begin{equation}
r\phi\left(\bar t,\boldsymbol{\bar x}\right)-\frac{\partial \phi}{\partial t}\left(\bar t,\boldsymbol{\bar x}\right)-\mathcal H_1 \left(\bar t,\boldsymbol{\bar x},D_{\boldsymbol{ x}}\phi\right)-\mathcal B_{b,\Delta}(\bar t,\boldsymbol{\bar x},\phi)-\mathcal B_{\Delta}(\bar t,\boldsymbol{\bar x},\phi)-\mathcal B_{w}(\bar t,\boldsymbol{\bar x},\phi)\leq 0,
\label{eq:subsub}
\end{equation}
holds. We define $\tau_1=\inf\big\{u>\bar{t}\,|\big(u,\,\boldsymbol{ X}_{\bar t,\,\boldsymbol{\bar x}}(u)\big)\notin B_{\epsilon}(\bar{t},\boldsymbol{\bar x})\big\}$ and we define the stopping time $\tau^*=\tau_1\wedge(\bar{t}+h)$. By the second part of the DP, there exist a control $\boldsymbol\nu^{*}\in\mathcal Z$ such that, for $\delta>0$, we have
\begin{equation*}
\begin{split}
V\big(\bar{t},\,&\boldsymbol{\bar x} \big)\leq\E\!\left[\int_{\bar{t}}^{\tau^{*}}\!\!\! \e^{-r(u-\bar t)}f_1\left(u,\boldsymbol{ X^{\nu^{*}}}_{\bar t,\,\boldsymbol{\bar x}}(u),{\nu^{*}}(u)\right)\!\rd u\!+\e^{-r(\tau^{*}-\bar{t})}\phi\left(\tau^{*}\!\!,\boldsymbol{ X^{\nu^{*}}}_{\bar t,\,\boldsymbol{\bar x}}(\tau^*)\right)  \right]+\delta\,\bar h.
\end{split}
\end{equation*}
We divide both sides by $-\bar \gamma$ and get 
\begin{equation*}
\begin{split}
-\delta\geq\frac{1}{\bar \gamma}\E\Bigg[\int_{\bar{t}}^{\tau^{*}}&-\Biggl(\e^{-r(u-\bar t)}f_1\left(u,\boldsymbol{ X^{\nu^{*}}}_{\bar t,\,\boldsymbol{\bar x}}(u),{\nu}(u)\right)-\e^{-r(u-\bar t)}\Biggl\{r\phi\left(u,\boldsymbol{ X^{\nu^{*}}}_{\bar t,\,\boldsymbol{\bar x}}(u)\right)\! \\
& - \frac{\partial \phi}{\partial t}\left(u,\boldsymbol{ X^{\nu^{*}}}_{\bar t,\,\boldsymbol{\bar x}}(u)\right)-\mathcal{\bar{H}}_1 \left(u,\boldsymbol{ X^{\nu^{*}}}_{\bar t,\,\boldsymbol{\bar x}}(t),D_{\boldsymbol{ x}}\phi,{\nu}(u)\right)- \mathcal B_{b,\Delta}\left(u,\boldsymbol{ X^{\nu^{*}}}_{\bar t,\,\boldsymbol{\bar x}}(u),\phi\right)\\
& -\mathcal B_{\Delta}\left(u,\boldsymbol{ X^{\nu^{*}}}_{\bar t,\,\boldsymbol{\bar x}}(u),\phi\right) -\bar{\mathcal B}_{w}\left(u,\boldsymbol{ X^{\nu^{*}}}_{\bar t,\,\boldsymbol{\bar x}}(u),\eta^{*}(u)\phi\right)\!\Biggr\}\Biggr)\rd u\Bigg].
\end{split}
\end{equation*}
For $h\rightarrow0$ and by the arbitrariness of $\delta$, (\ref{eq:subsub}) follows.$\hfill\square$

\begin{prop}
{\itshape Let $U$ (resp. $V$) be a viscosity subsolution (resp. supersolution) of  (\ref{eq:pde1}). If $U(T,\boldsymbol{x})\leq V(T,\boldsymbol{x})$ on $\mathcal O$, then $U\leq V$ on $[0,T]\times\mathcal O$.
\label{eq:dpuniq}}
\end{prop}
\noindent{\itshape Proof.} Let $U$ be a subsolution and $V$ be a supersolution. Since $U$, $V\in PB$, then there exist a $p>1$ such that
\begin{equation}
\frac{|U(t,\boldsymbol{ x})|+|V(t,\boldsymbol{ x})|}{\left(1+\|\boldsymbol{ x}\|_{p}^{p}\right)}<\infty,
\label{eq:polynomial}
\end{equation}
where the operator $\|\cdot\|_{p}^{p}$ is the $L_{p}$-norm raised to the $p$-th power. Let $\tilde V^{\epsilon}(t,\boldsymbol{ x}):=V(t,\boldsymbol{ x})+\epsilon \kappa(t,\boldsymbol{ x})$, where $\epsilon>0$ and $\kappa(t,\boldsymbol{ x})=\e^{-\zeta t}\big(1+\|\boldsymbol{ x}\|_{2p}^{2p}\big)$, for $\zeta>0$. Then $\tilde V^{\epsilon}$ is a supersolution of (\ref{eq:pde1}). Indeed, let $\phi(t,\boldsymbol{ x})$ be the test function for $\tilde V^{\epsilon}$, then the test function for $V$ is  $\phi(t,\boldsymbol{ x})-\epsilon\kappa(t,\boldsymbol{ x})$. First note that, due to assumptions (\ref{eq:hhh})-(\ref{eq:hhh3}) and the fact that $\mathcal Z$ is a compact set, we have
\begin{equation*}
\begin{split}
r\epsilon\kappa\left(t, \boldsymbol{ x}\right)-\frac{\partial \epsilon\kappa}{\partial t}\left( t,\boldsymbol{ x}\right)-\sup_{{ \boldsymbol{v}}\in\mathcal Z}\mathcal{\bar H}_1 \left(t,\boldsymbol{ x},D_{\boldsymbol{ x}}\epsilon\kappa,\boldsymbol{v}\right)-\mathcal B_{b,\Delta}( t,\boldsymbol{ x},\epsilon\kappa)&\\
-\mathcal B_{\Delta}( t,\boldsymbol{ x},\epsilon\kappa)-\mathcal B_{w}( t,\boldsymbol{ x},\epsilon\kappa)&\geq 0,
\end{split}
\end{equation*}
for $\zeta$ sufficiently large. By the supersolution property of $V$, we have
\begin{equation*}
\begin{split}
&r(\phi-\epsilon\kappa)\left(t, \boldsymbol{ x}\right)-\frac{\partial (\phi-\epsilon\kappa)}{\partial t}\left( t,\boldsymbol{ x}\right)-\mathcal H_1 \left(t,\boldsymbol{ x},D_{\boldsymbol{ x}}\left(\phi-\epsilon\kappa\right)\right)-\mathcal B_{b,\Delta}( t,\boldsymbol{ x},\phi-\epsilon\kappa)\\
&-\mathcal B_{\Delta}( t,\boldsymbol{ x},\phi-\epsilon\kappa)-\mathcal B_{w}( t,\boldsymbol{ x},\phi-\epsilon\kappa)\geq 0,
\end{split}
\end{equation*}
and recalling that $\sup(A+B)\leq\sup A+\sup B$, we have
\begin{equation*}
\begin{split}
r\phi\left(t, \boldsymbol{ x}\right)-\frac{\partial \phi}{\partial t}\left( t,\boldsymbol{ x}\right)-\mathcal H_1 \left(t,\boldsymbol{ x},D_{\boldsymbol{ x}}\phi\right)-\mathcal B_{b}( t,\boldsymbol{ x},\phi)-\mathcal B_{b,\Delta}( t,\boldsymbol{ x},\phi)-\mathcal B_{w}( t,\boldsymbol{ x},\phi)&\geq \\
r\epsilon\kappa\left(t, \boldsymbol{ x}\right)-\frac{\partial \epsilon\kappa}{\partial t}\left( t,\boldsymbol{ x}\right)-\sup_{\boldsymbol{ v}\in\mathcal Z}\mathcal{\bar H}_1 \left(t,\boldsymbol{ x},D_{\boldsymbol{ x}}\epsilon\kappa,\boldsymbol{v}\right)-\mathcal B_{b,\Delta}( t,\boldsymbol{ x},\epsilon\kappa)&\\
-\mathcal B_{\Delta}( t,\boldsymbol{ x},\epsilon\kappa)-\mathcal B_{w}( t,\boldsymbol{ x},\epsilon\kappa)&\geq 0.
\end{split}
\end{equation*}
Since by (\ref{eq:polynomial}) $\lim_{\boldsymbol{ x}\rightarrow\infty}\sup_{[0,T]}(U-\tilde V^{\epsilon})(t,\boldsymbol{ x})=-\infty$, we can assume w.l.o.g. that 
$$
\mathcal M:=\max_{[0,T]\times\mathcal O}\left(U\left(t,\boldsymbol{ x}\right)-V\left(t,\boldsymbol{ x}\right)\right),
$$
is attained at  $(\bar t,\boldsymbol{\bar{ x}})\in[0,T]\times\Sigma$, where $\Sigma\subset\mathcal O$ is a compact set. In order to prove Proposition $\ref{eq:dpuniq}$, it suffices to show that $\mathcal M<0$. Suppose by contradiction that there exists a $(\bar t,\boldsymbol{ \bar x})\in[0,T)\times\Sigma$ such that $\mathcal M>0$. For $\epsilon>0$, we define the function $\Psi^{\epsilon}$ by
$$
\Psi^{\epsilon}\left(t_{1},t_{2},\boldsymbol{x}_{1},\boldsymbol{x}_{2}\right)=U\left(t_{1},\boldsymbol{x}_{1}\right)-V\left(t_{2},\boldsymbol{x}_{2}\right)-\psi^{\epsilon}\left(t_{1},t_{2},\boldsymbol{x}_{1},\boldsymbol{x}_{2}\right),
$$
where 
$$
\psi^{\epsilon}\left(t_{1},t_{2},\boldsymbol{x}_{1},\boldsymbol{x}_{2}\right):=\frac{1}{2\epsilon}\left(|t_{1}-t_{2}|^{2}+\|\boldsymbol{x}_{1}-\boldsymbol{x}_{2}\|_{2}^{2}\right).
$$
The function $\Psi^{\epsilon}$ is continuous and admits a maximum point $\mathcal M^{\epsilon}$, where $\mathcal M\leq \mathcal M^{\epsilon}$, at $m=\big(t^\epsilon_{1},t_2^\epsilon,\boldsymbol{ x}_1^\epsilon,\boldsymbol{ x}_2^\epsilon\big)$. That is, the function $U(t_{1},\boldsymbol{ x}_1)-\psi^\epsilon(t_{1},t_{2},\boldsymbol{ x}_1,\boldsymbol{ x}_2)$ has its maximum at $m$ and $
V(t_{2},\boldsymbol{ x}_1)-(-\psi^\epsilon(t_{1},t_{2},\boldsymbol{ x}_1,\boldsymbol{ x}_2))
$ has its minimum at $m$. Also,  formally $\lim_{\epsilon \rightarrow 0}\big({t^\epsilon_1},{t^\epsilon_2},{\boldsymbol{ x}^\epsilon_1},{\boldsymbol{ x}^\epsilon_2}\big)$ converges, up to a subsequence, to $(\bar{t},\bar{t},\bar{\boldsymbol{ x}},\bar{\boldsymbol{ x}})$ (see Crandall et al. \cite{11} for details). We let $o^{\epsilon}=(t_{1}-t_{2})/\epsilon$, and define the vector $\boldsymbol{p}^{\epsilon}$  by
$$
\boldsymbol{ p}^{\epsilon}=\left(p^{\epsilon}_{1},p^{\epsilon}_{2},p^{\epsilon}_{3},p^{4}\right)=\left(\frac{1}{\epsilon}\big(x^{\epsilon}_{1}-x^{\epsilon}_{2}\big),\frac{1}{\epsilon}\big(s^{b,\epsilon}_{1}-s^{b,\epsilon}_{2}\big),\frac{1}{\epsilon}\big(\Delta^{\epsilon}_{1}-\Delta^{\epsilon}_{2}\big),\frac{1}{\epsilon}\big(w^{\epsilon}_{1}-w^{\epsilon}_{2}\big)\right).
$$
We can apply the viscosity subsolution and supersolution properties at the point $m$ to see that, for  $\delta\in[0,1]$ and for some $\varphi_{1}$, $\varphi_{2}\in \mathcal C^{1,1}([0,T]\times\mathcal O)$ such that $U-\varphi_{1}$ ($V-\varphi_{2}$) has its maximum (minumum) at $t^{\epsilon}_{1},\boldsymbol{x}^{\epsilon}_{1}\rightarrow\bar t,\bar{\boldsymbol{x}}$ $\big(t^{\epsilon}_{2},\boldsymbol{x}^{\epsilon}_{2}\rightarrow\bar t,\bar{\boldsymbol{x}}\big)$, we have
\begin{equation*}
\begin{split}
&rU\left({t_{1}^\epsilon},{\boldsymbol{ x}^\epsilon_1}\right)-o^{\epsilon}-{\mathcal H}_1\left(t^\epsilon_{1},\boldsymbol{ x}_1^\epsilon,\boldsymbol{ p}^{\epsilon}\right)-\mathcal B^{\delta}_{b,\Delta}\left(t^\epsilon_{1},\boldsymbol{ x}_1^\epsilon,p^{\epsilon}_{2},U\right)-\mathcal B_{\delta}^{b,\Delta}\left(t^\epsilon_{1},\boldsymbol{ x}_1^\epsilon,\varphi_{1}\right)\\
&-\mathcal B^{\delta}_{\Delta}\left(t^\epsilon_{1},\boldsymbol{ x}_1^\epsilon,p^{\epsilon}_{3},U\right)-\mathcal B_{\delta}^{\Delta}\left(t^\epsilon_{1},\boldsymbol{ x}_1^\epsilon,\varphi_{1}\right)-\mathcal B_{w}\left(t^\epsilon_{1},\boldsymbol{ x}_1^\epsilon,U\right)\leq 0
\end{split}
\end{equation*}
and
\begin{equation*}
\begin{split}
&rV\left({t_{2}^\epsilon},{\boldsymbol{ x}^\epsilon_2}\right)-o^{\epsilon}-{\mathcal H}_1\left(t^\epsilon_{2},\boldsymbol{ x}_2^\epsilon,\boldsymbol{ p}^{\epsilon}\right)-\mathcal B^{\delta}_{b,\Delta}\left(t^\epsilon_{2},\boldsymbol{ x}_2^\epsilon,p^{\epsilon}_{2},V\right)-\mathcal B_{\delta}^{b,\Delta}\left(t^\epsilon_{2},\boldsymbol{ x}_2^\epsilon,\varphi_{2}\right)\\
&-\mathcal B^{\delta}_{{\Delta}}\left(t^\epsilon_{2},\boldsymbol{ x}_2^\epsilon,p^{\epsilon}_{3},V\right)-\mathcal B_{\delta}^{{\Delta}}\left(t^\epsilon_{2},\boldsymbol{ x}_2^\epsilon,\varphi_{2}\right)-\mathcal B_{w}\left(t^\epsilon_{2},\boldsymbol{ x}_2^\epsilon,V\right)\geq 0.
\end{split}
\end{equation*}
We can subtract the two inequalities and take the limit for $\epsilon,\delta \rightarrow 0$ to get $
r\left[U\left(\bar{t},\bar{\boldsymbol{ x}}\right)-V\left(\bar{t},\bar{\boldsymbol{ x}}\right)\right]\leq0,
$ which concludes  the proof.$\hfill\square$

\section*{Acknowledgments}
The authors thank A. Cartea, J. Sekine and J. Walton for useful discussions. A. Macrina acknowledges support from the African Collaboration for Quantitative Finance and Risk Research (ACQuFRR), University of Cape Town.

\end{document}